\documentclass[a4paper,11pt]{article}
\pdfoutput=1 

\usepackage{jheppub} 

\usepackage[T1]{fontenc} 

\pdfoutput=1
\usepackage{graphicx}
\usepackage[usenames,dvipsnames]{color}

\usepackage{dcolumn}
\usepackage{bm}

\usepackage{hyperref}
\hypersetup{colorlinks,citecolor= blue,linkcolor= blue, urlcolor=blue}

\usepackage{amsmath}
\usepackage{bbm}
\usepackage{amsfonts}
\usepackage{amssymb}
\usepackage{latexsym}
\usepackage{graphicx}
\usepackage[english]{babel}
\usepackage{multirow}
\usepackage{float}
\usepackage{url}
\usepackage{slashed}
\usepackage{xcolor} 
\usepackage{array}
\usepackage{cases}

\newcommand{\be}{\begin{equation}}
\newcommand{\ee}{\end{equation}}
\newcommand{\ba}{\begin{array}}
\newcommand{\ea}{\end{array}}
\newcommand{\bea}{\begin{eqnarray}}
\newcommand{\eea}{\end{eqnarray}}

\usepackage{ulem,fancyvrb}
\usepackage{xcolor}

\title{Millicharged particles from proton bremsstrahlung in the atmosphere}

\author[a,b]{Mingxuan Du,}
\author[a]{Rundong Fang}
\author[a,c]{and Zuowei Liu}

\affiliation[a]{Department of Physics, Nanjing University, Nanjing 210093, China}
\affiliation[b]{Center for High Energy Physics, Peking University, Beijing 100871, China}
\affiliation[c]{CAS Center for Excellence in Particle Physics, Beijing 100049, China}

\emailAdd{mg1722004@smail.nju.edu.cn}
\emailAdd{141150012@smail.nju.edu.cn}
\emailAdd{zuoweiliu@nju.edu.cn}

\abstract{

Light millicharged particles can be copiously produced from meson decays in cosmic ray collisions with the atmosphere, leading to detectable signals in large underground neutrino detectors. In this paper we study a new channel for generating atmospheric millicharged particles, the proton bremsstrahlung process. We find that the proton bremsstrahlung process leads to a significantly higher flux of millicharged particles compared to meson decays and, for certain masses, results in a one-order-of-magnitude improvement in the flux. Consequently, Super-K constraints on $\varepsilon^2$ for sub-GeV MCPs are improved by half order of magnitude. We further note that the study on the proton bremsstrahlung process can be extended to a variety of new physics particle searches in atmospheric collisions and in low energy proton accelerators.

}

\begin{document}

\maketitle
\flushbottom

\section{Introduction}

Although the standard model (SM) has achieved remarkable 
success in describing the microscopic world, 
there is overwhelming evidence pointing to the 
existence of new physics beyond the standard model (BSM). 
Recently, a great amount of effort has been devoted 
to BSM models with a dark sector (or hidden sector)
that only has portal interactions with the SM sector 
\cite{Alexander:2016aln}. 
Light dark sector particles 
with mass in the MeV-GeV range 
can still 
have a somewhat sizable coupling 
to the SM sector, 
despite the great progress both in the 
energy and intensity frontiers
\cite{Lanfranchi:2020crw, Agrawal:2021dbo}.
One class of such new particles is called 
millicharged particles (MCPs), which 
interact with the SM photon 
via a small electric charge 
\cite{Jaeckel:2010ni, Fabbrichesi:2020wbt}.

MCPs can naturally arise in models with a  
``hypercharge portal'', which can be materialized 
either in a kinetic mixing portal 
\cite{Holdom:1985ag, Holdom:1986eq, Foot:1991kb}, 
or in a Stueckelberg portal  
\cite{Kors:2004dx, Cheung:2007ut, Feldman:2007wj}. 
Recently, searches for MCPs have been pursued 
extensively 
both in astrophysics and cosmology 
and in accelerator experiments on Earth. 
For light MCPs with mass $\lesssim$ MeV, 
the stringent constraints mainly come from 
astrophysics and cosmology processes, including
stellar evolution  
\cite{Dobroliubov:1989mr, Mohapatra:1990vq, Davidson:1991si, Vinyoles:2015khy, Davidson:1993sj, Vogel:2013raa, Davidson:2000hf, Chang:2018rso} and  
big bang nucleosynthesis
\cite{Davidson:1991si, Davidson:1993sj, Davidson:2000hf, Vogel:2013raa, Vinyoles:2015khy}.  
Constraints on MCPs with mass above MeV 
are mainly dominated by
accelerator experiments, including  
fixed target experiments with 
an electron beam 
\cite{Prinz:1998ua, Golowich:1986tj, Soper:2014ska, Berlin:2018bsc, Gninenko:2018ter, Chu:2018qrm, Anchordoqui:2021ghd}  
and with a proton beam
\cite{Magill:2018tbb, Harnik:2019zee, ArgoNeuT:2019ckq},
and collider experiments 
\cite{Davidson:1991si, Davidson:2000hf, CMS:2012xi, Haas:2014dda, Ball:2020dnx}. 
Recently, a number of new search 
strategies for MCPs have been proposed 
\cite{Ball:2016zrp, Liu:2018jdi, Kelly:2018brz, Liu:2019ogn, Foroughi-Abari:2020qar, Liang:2019zkb,  Afek:2020lek, Kim:2021eix, Gorbunov:2021jog, Carney:2021irt, Budker:2021quh, Gorbunov:2022bzi, Kling:2022ykt, Gorbunov:2022dgw}.
We note that dark matter (DM) can be millicharged 
\cite{Brahm:1989jh, Cheung:2007ut, Feldman:2007wj, Feng:2009mn, Cline:2012is, Foot:2014uba}, 
which may provide an interpretation to 
the recent EDGES $21$ cm anomaly  
\cite{Bowman:2018yin,Barkana:2018qrx, Munoz:2018pzp, Berlin:2018sjs, Barkana:2018lgd, Boddy:2018wzy, Fialkov:2018xre, Kovetz:2018zan, Slatyer:2018aqg, Liu:2019knx, Creque-Sarbinowski:2019mcm, Aboubrahim:2021ohe, Li:2021kso}.

\begin{figure}[htbp]
\begin{centering}
\includegraphics[width= 0.5 \columnwidth]{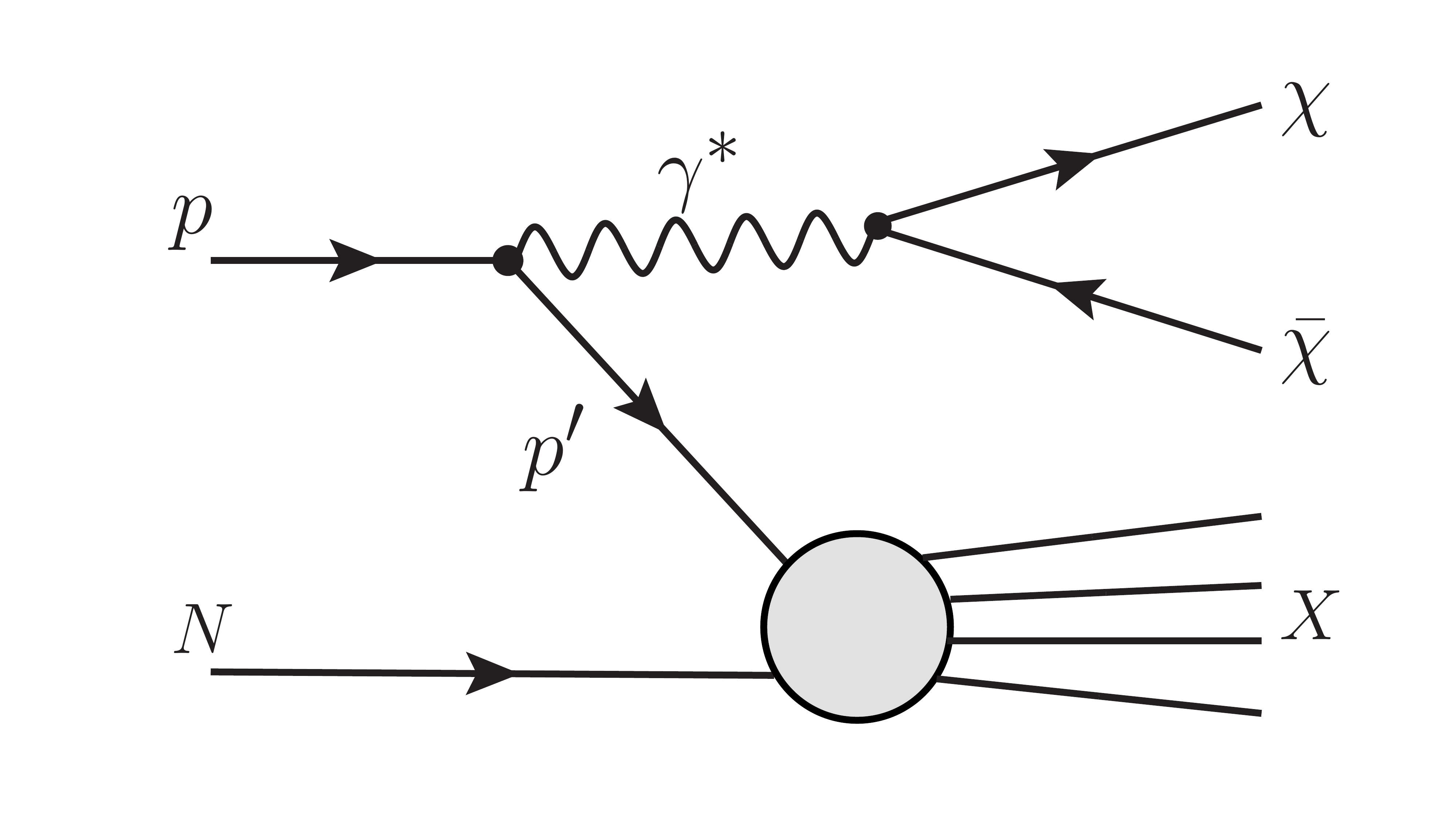}
\caption{Feynman diagram to produce MCP $\chi$ 
in the proton bremsstrahlung process in  
collisions between the cosmic proton $p$ and the  nucleus $N$ in the atmosphere.}
\label{fig:pb-diagram}
\end{centering}
\end{figure}

One interesting probe of MCPs, 
which has been studied recently \cite{Plestid:2020kdm, Kachelriess:2021man, ArguellesDelgado:2021lek}, 
is through 
collisions between cosmic rays and the 
atmosphere \cite{Alvey:2019zaa}.\footnote{For atmospheric production of other new physics particles, see e.g.,
Refs.~\cite{Kusenko:2004qc, Ando:2007ds, Yin:2009yt, Asaka:2012hc, Bueno:2013mfa, Masip:2014xna, Arguelles:2019ziu, Coloma:2019htx, Meighen-Berger:2020eun, Su:2020zny, Candia:2021bsl, Iguro:2021xsu, Arguelles:2022fqq, Darme:2022bew, Cheung:2022umw, Su:2022wpj, PandaX:2023tfq, Fischer:2023bfn}.} 
Large underground neutrino detectors
such as Super-K are the ideal experiments 
to detect 
atmospheric MCPs, 
which can lead to competitive limits 
\cite{Plestid:2020kdm}. 
However, only the MCP flux 
from meson decays (MD) 
in the atmospheric air shower 
has been considered in 
Refs.~\cite{Plestid:2020kdm, Kachelriess:2021man, ArguellesDelgado:2021lek}. 
In this paper we consider another important channel 
to produce MCPs in the atmosphere, 
the proton bremsstrahlung (PB) process.
In the PB process, a cosmic proton radiates a pair of MCPs
when it collides with nuclei in the atmosphere;
the Feynman diagram of this process
is shown in Fig.~\ref{fig:pb-diagram}.

The PB process is often computed in the 
Fermi-Weizs\"acker-Williams (FWW) approximation
\cite{Fermi:1924, Williams:1934, Weizsacker:1934};
see e.g., Refs.~\cite{Blumlein:2013cua, deNiverville:2016rqh, Tsai:2019buq,  Feng:2017uoz, Foroughi-Abari:2021zbm, Du:2021cmt} 
for recent calculations of the PB process 
in accelerators, which can lead to competitive 
contributions to dark photons as  
the MD process. {However}, 
the FWW approximation requires relativistic and 
collinear conditions for initial and final state 
particles, which may not be satisfied in 
atmospheric collisions. 
To take into account the contributions from 
low energy protons, 
which are the dominant component of the cosmic rays, 
we develop a new method 
to compute the PB process.
We find that the dominant production channel 
for the atmospheric sub-GeV MCPs  
is the PB process, 
rather than the previously studied MD process. 
Consequently, the Super-K constraints 
on $\varepsilon^2$ for sub-GeV MCPs 
are improved by 
about half order of magnitude,
when the PB process is taken into account. We also investigate MCP limit from the Hyper-K detector, 
the successor of Super-K.

\section{Atmospheric MCP flux}

To study atmospheric MCPs,
we consider a low-energy phenomenological model 
such that
the MCP is the only 
new particle in the dark sector, 
which is usually assumed in various 
terrestrial studies on MCPs.
The interaction Lagrangian between 
the MCP $\chi$ and the SM photon $A_\mu$ is 
$\mathcal{L}_{\rm int} =  
\varepsilon e A_\mu \bar{\chi} \gamma^\mu \chi$,
where  
$e$ is the QED coupling constant, 
and $\varepsilon \ll 1$ is the millicharge.

To compute the MCP flux at the surface of the Earth,  
we use the one-dimensional approximation 
(see e.g., Refs.~\cite{Gondolo:1995fq, Giunti:2007ry, Lipari:2000wu}) 
such that the final state particles produced 
in proton-air collisions are assumed to 
have the same direction as the incident proton. 
We further neglect the small attenuation 
effects of the atmosphere on MCPs.
Cosmic protons, however, interact so strongly with
the atmosphere that they are effectively absorbed
before reaching the ground. 
This leads to an isotropic 
MCP flux on the Earth's surface 
for the zenith angle $\theta_s < \pi/2$. 
Thus, in this analysis, 
we will only 
compute the MCP flux on the surface of Earth with 
the zenith angle $\theta_s =0$ 
(due to collisions from protons that have 
a vertically downward momentum, 
in the one-dimensional approximation), 
which is then taken to be the flux for all 
directions with $\theta_s < \pi/2$.

The MCP flux on the Earth's surface 
for the zenith angle $\theta_s =0$ 
can be obtained by the cascade equation \cite{Gondolo:1995fq} 
\be 
\frac{d^2 \Phi_\chi^s}{d E_\chi^s 
d \Omega_\chi^s} = 
\iint
d h  d E_p \frac{d^2 \Phi_p(h)}{d E_p d \Omega_p} 
n_{T}(h) \sigma_{pT}
\sum_i \frac{d N^i_\chi}{d E_\chi^{s}},
\label{eq: ground chi flux}
\ee
where $h$ is the height,
$E_p$ is the proton energy, 
$\Phi_\chi^s$ is the MCP flux at the surface of Earth, 
$\Phi_p(h)$ is the proton flux at height $h$, 
$n_{T}(h)$ is the number density of air at height $h$, 
$\sigma_{pT}$ is the in-elastic proton-air cross section, 
${d N^i_\chi}/{d E_\chi^s}$ is the energy spectrum 
of $\chi$ per proton-air interaction in the production 
channel denoted by $i$, which can be either the 
PB channel or the 
MD channel.
We present our MD calculations and comparisons with
Refs.~\cite{Plestid:2020kdm, Kachelriess:2021man, ArguellesDelgado:2021lek}
in Appendix \ref{appendix:MD}.

In our analysis we only consider proton-nitrogen collisions 
in the atmosphere,  
and adopt a constant $\sigma_{pT} \simeq 253$ mb  
\cite{Pierog:2013ria, Ulrich:crmc},
since the cross section has a weak dependence on the 
proton energy.
We use the NRLMSISE-00 atmosphere model \cite{picone2002nrlmsise} 
for the air density $n_T(h)$. 
The cosmic ray proton flux at the top
of the atmosphere can be well approximated by a 
power law \cite{ParticleDataGroup:2018ovx}:
\be
\frac{d^2 \Phi_p}{d E_p d \Omega_p }(h_{\max}) =  
\frac{0.74\times 1.8 \times 10^{4}}{\rm{m^2\ s\ sr\ GeV}}
\left(\frac{E_p}{\rm{GeV}}\right)^{-2.7}, 
\label{eq: cosmic proton spectrum at top}
\ee
where $h_{\max}=65$ km.
We obtain the proton flux at height $h$
by using the cascade equation \cite{Gondolo:1995fq}
\be
\frac{d}{d h} 
\left[ \frac{d^2 \Phi_p}{d E_p d \Omega_p}(h) \right]
= \sigma_{pT} n_{T}(h)
\frac{d^2 \Phi_p}{d E_p d \Omega_p}(h). 
\ee 
We also carry out analyses by using the 
actual cosmic ray data \cite{ParticleDataGroup:2018ovx} 
(see Appendix \ref{appendix: our FWW compare}),
which agree with analyses using the power-law spectrum.

\section{MCPs in the PB channel}

The energy spectrum of MCPs 
in the PB process can be computed by \cite{Gninenko:2018ter}
\begin{align}
     \frac{d N_\chi^{\rm PB}}
     {d E_\chi} & = 
  \frac{\varepsilon^2 e^2}{6 \pi^2}
\int \frac{d k^2}{k^2} 
\sqrt{1 - \frac{4 m_\chi^2}{k^2}}
\left( 1 + \frac{2 m_\chi^2}{k^2} \right) 
\nonumber \\ \times & 
\int d E_k 
\frac{1}{\sigma_{pT}}
\frac{d \sigma_{\rm PB}}{d E_k} 
\frac{\Theta\left(E_\chi- E_{-}\right) \Theta\left(E_{+} - E_\chi \right)}{E_{+} - E_{-}} ,
\label{eq:PB:MCP:spectra}
\end{align}
where $m_\chi$ is the mass of the MCP. 
{Here} ${d \sigma_{\rm PB}}/{d E_{k}}$ 
is the cross section of the $pN\to \gamma^* X$ process 
with the following sum rule
for the off-shell photon $\gamma^*$ \cite{Liang:2021kgw}
\begin{equation}
\sum_i \epsilon_{\mu}^{i} (k)\epsilon_{\nu}^{i *} (k) 
= - g_{\mu\nu} + \frac{k_{\mu} k_{\nu}}{k^2},
\end{equation}
{where}
$k^\mu=(E_k,\vec k)$ {is} the photon momentum, 
and $X$ denotes the final state particles 
in addition to the off-shell photon $\gamma^*$.
The maximal and minimum 
energies of MCPs are given by 
$E_{\pm} \equiv \gamma (E_{\chi}^r 
\pm  \beta p_{\chi}^r)$ 
where 
$\gamma = (1-\beta^2)^{-1/2}
= E_{\gamma^*}/\sqrt{k^2}$, 
and 
$E_\chi^r$ ($p_\chi^r$) is 
the energy (magnitude of momentum)
of MCPs in the rest frame of the off-shell photon.

The cross section ${d \sigma_{\rm PB}}/{d E_{k}}$ 
in the PB process $pN\to \gamma^* X$ 
is often computed in 
the FWW approximation 
\cite{Fermi:1924, Williams:1934, Weizsacker:1934},
in which relativistic and  
collinear conditions 
are assumed for the protons and the photon
\cite{Kim:1973he, Tsai:1973py, Blumlein:2013cua, deNiverville:2016rqh, Feng:2017uoz, Tsai:2019buq, Foroughi-Abari:2021zbm, Du:2021cmt}.
These conditions are 
excellent approximations 
to physics in 
accelerators with an energetic proton beam. 
However, for the cosmic rays, the 
low-energy protons are the dominant component 
of the cosmic flux, as indicated by 
the power law of $E^{-2.7}$.
In an attempt to take into account 
the low-energy protons, we develop
a new method to compute the PB process.
We compare our method with the FWW approximation 
in Appendix \ref{appendix: our FWW compare}.

In our method, we first compute the splitting kernel for 
the $p \to \gamma^* p$ process, 
by taking {the} ratio of the differential 
cross section of the 2-to-3 process, 
$p (p_1) \bar{p} (p_2) \to 
 p (p_3) \bar{p} (p_4) \gamma^* (k)$, 
where the momentum for each particle is 
given inside the parenthesis,
to the total cross section of the 
corresponding 2-to-2 process, 
$p \bar{p} \to p \bar{p} $. 
Both processes are mediated by an s-channel photon. 
Thus one has\footnote{See also Ref.~\cite{Foroughi-Abari:2021zbm}
for a similar treatment, but
for t-channel pomeron exchange diagrams
in relativistic limits.}
\begin{equation}
    \frac{d^2 \mathcal{P}_{p \to \gamma^* p}}
{d E_k d \cos\theta_k} =
\frac{1}{\sigma(p \bar{p} \to p \bar{p} )}
\frac{d^2\sigma(p \bar{p} \to \gamma^* p \bar{p})}
{d E_k \cos\theta_k}, 
\end{equation}
where the 2-to-3 process is evaluated at $s = (p_1 + p_2)^2$, and
$\sigma(p \bar{p} \to p \bar{p} )$ 
is evaluated at $s_k=(p_3+p_4)^2$.
We first evaluate the splitting kernel in the center-of-mass (CM) frame, 
\begin{equation}
  \frac{d^2 \mathcal{P}_{p \to \gamma^* p}}
{d E^0_{k} d \cos\theta_{k}^0} = 
\frac{1}{\sigma_{2\to 2} (s_k)}
\frac{\int d E^0_{3} \int d \phi_{3,k}^0
\overline{ \left| \mathcal{M}_{2 \to 3} \right| }^2} {512 \pi^4 E_1^0 E_{2}^0 
\left|\vec{v}_1^{\, 0} -\vec{v}_{2}^{\, 0} \right|}, 
\label{eq:diff_xsection_cm}  
\end{equation}
where $\mathcal{M}_{2\to3}$ is the matrix element
of the $p \bar{p} \to \gamma^* p \bar{p}$  process, 
${\sigma_{2\to 2} (s_k)}$ is the cross section of the 
$p \bar{p} \to p \bar{p}$  process, 
$E^0_{3}$ is the energy of the final state proton,
$\phi_{3,k}^0$ is the azimuth angle of $\vec p_3$ 
in the transverse plane of $\vec k$, 
and $\theta_{k}^0$ is the angle
between $\gamma^*$ and the initial state proton.
We only consider the initial radiation diagrams  
in computing the $\mathcal{M}_{2\to3}$ matrix element 
so that the splitting kernel is compatible with 
the process shown in Fig.~(\ref{fig:pb-diagram}).
We find that 
\footnote{We note that 
in the limit where both the proton and the photon 
are massless, 
the splitting kernel given in 
Eq.~\eqref{eq:SK:analytic} reduces to the 
well-known NLO radiator \cite{Nicrosini:1989pn,Montagna:1995wp}: 
\begin{equation}
\frac{d^2 \mathcal{P}_{p \to \gamma^* p}}{d E_k^0 d \cos\theta_{k}^0}  
= \frac{1}{E_p^0}\frac{\alpha}{\pi} \frac{1}{x_{\gamma}}
    \left[\frac{x_\gamma^2-2x_\gamma+2}{1- (\cos\theta_{k}^0)^2}-{x_\gamma^2 \over 2}\right], 
\end{equation}
where $x_\gamma = E_k^0/E_p^0$ with 
$E_p^0$ being the proton energy in the CM frame.}
\begin{equation}
\frac{d^2 \mathcal{P}_{p \to \gamma^* p}}{d E_k^0 d \cos\theta_{k}^0}  
= \frac{2\alpha}{\pi E_k^0} 
 \frac{2\beta_f \beta_k}{(3-\beta_f^2)\beta_i}
\frac{N}{s\left[x^2-(1-y)^2\right]^2},
\label{eq:SK:analytic}
\end{equation}
where 
$s_k=s+m_k^2-2 E_k^0 \sqrt{s}$,
$m_k = \sqrt{k^2}$,
$\beta_i=\sqrt{1-4m_p^2/s}$, 
$\beta_f=\sqrt{1-4m_p^2/s_k}$, 
$\beta_k=\sqrt{1-m_k^2/(E_k^0)^2}$,
$x=\beta_i \beta_k \cos\theta_{k}^0$,
$y= m_k^2/(E_k^0\sqrt{s})$, 
and 
\begin{align}
N = &  
(1-y)^2
\left[ \beta_i^2 (2 m_p^2+s_k+m_k^2) + 
(E_k^0)^2 \beta_k^2 \right]
\nonumber \\ & 
- \left[ 2 m_p^2  + m_k^2(1+s_k/s)+ s_k \right]x^2 
- (E_k^0)^2 x^4.
\end{align}
We then use the resulting splitting kernel 
to compute the differential cross-section ${d \sigma_{\rm PB}}/{d E_{k}}$ of the 
$pN\to \gamma^* X$ process 
in the lab frame:  
\be
\frac{d \sigma_{\rm PB}}{d E_{k}} =
\int d \cos\theta_{k}
\left|\frac{\vec{k}}{\vec{k}^0}\right| \left| F_V (k) \right|^2
\left|F_* \left(p_p-k\right)\right|^2
\frac{d^2 \mathcal{P}_{p \to \gamma^* p}}{d E^0_{k} d \cos\theta_{k}^0}
\sigma_{pT}(s_2')
\label{eq:dsigmadEk:lab:frame}
\ee
where $\sigma_{pT}$ 
is the proton-nitrogen cross section 
evaluated at $s'_2=\left(p_p + p_N - k \right)^2$ 
with $p_p$ ($p_N$, $k$)  
being the momentum of the proton (nitrogen, photon), 
and the $|\vec{k}|/|\vec{k}^0|$ factor is the Jacobian 
between $\left(E^0_{k}, \cos\theta_{k}^0\right)$ in the CM frame 
and $(E_k, \cos\theta_k)$ in the lab frame 
with $\vec{k}$ ($\vec{k}^0$) being the 3-momentum
of the off-shell photon in the lab (CM) frame.
The upper and lower bound of the integrate region is $\cos\theta^{\rm min}_k = {\rm Max}[-1, (\gamma_i E_k 
 - E^{0, \rm max}_k)/\gamma_i \beta_i |\vec{k}|]$
and $\cos\theta^{\rm max}_k = {\rm Min}[1, (\gamma_i E_k 
 - \sqrt{k^2})/\gamma_i \beta_i |\vec{k}|]$,
 where $E^{0, \rm max}_k =(4 (E^0_p)^2 - 4 m^2_p + k^2)/{ 4 E^0_p}$
is the maximal energy of the off-shell photon in the CM frame and $\gamma_i = (1 - \beta_i^2)^{-1/2}$,
and $E^0_p$ is the energy of the proton in the CM frame.
In Eq.~\eqref{eq:dsigmadEk:lab:frame}, 
we have included two form factor functions: 
the vector meson form factor $F_V$, and 
the form factor $F_*$ that accounts for the off-shell-ness of the 
intermediate proton.

The vector meson form factor $F_V$ is included
for the $p \gamma^* p$ vertex, 
and is
the time-like form factor due to 
several light vector mesons 
\cite{Faessler:2009tn, deNiverville:2016rqh, Feng:2017uoz, Foroughi-Abari:2021zbm, Du:2021cmt} 
\be
F_V(k) = \sum_{V =\rho \, \rho' \, \rho'' \, \omega \, \omega' \, \omega''} \frac{f_V m_V^2}{m_V^2 - k^2 - i m_V \Gamma_V}, 
\label{eq:PB_formfactor}
\ee
where $m_V$ ($\Gamma_V$) is the mass (decay width) of the vector meson, 
$m_{\rho} = m_{\omega} = 0.77$ GeV,
$m_{\rho'} = m_{\omega'} = 1.25$ GeV,
$m_{\rho''} = m_{\omega''} = 1.45$ GeV,
$\Gamma_{\rho} = 0.150$ GeV,
$\Gamma_{\omega} = 0.0085$ GeV,
$\Gamma_{\rho'} = \Gamma_{\omega'} = 0.3$ GeV,
$\Gamma_{\rho''} = \Gamma_{\omega''} = 0.5$ GeV,
$f_{\rho} = 0.616$, $f_{\rho'} = 0.223$, $f_{\rho''} = -0.339$, $f_{\omega} = 1.011$, $f_{\omega'} = -0.881$, and $f_{\omega''} = 0.369$.
Because the time-like form factor 
is in the kinematic region $0<q^2<4m_p^2$,  
which has not been probed 
directly in experiments with proton beams, 
the expressions in Eq.~\eqref{eq:PB_formfactor}
are obtained from 
a generalized vector meson dominance (VMD) model 
by assuming both protons in the vertex to be on-shell. 
To account for the effects due to the 
off-shell-ness of the intermediate proton, 
following Ref.~\cite{Feuster:1998cj, Foroughi-Abari:2021zbm}, 
we include the following form factor: 
\be
F_* (p') = \frac{\Lambda^4}{ \Lambda^4 + (p'^2 - m_p^2)^2},
\label{eq: off-shell FF}
\ee
where 
$m_p$ is the proton mass, 
$p'$ is the four-momentum of the intermediate proton. 
Ref.~\cite{Foroughi-Abari:2021zbm} considered the range 
1 GeV $<\Lambda<$ 2 GeV. 
We find that changes in $\Lambda$ within this range can cause the atmospheric MCP production from the PB process to fluctuate by about one order of magnitude. 
In our analysis we adopt 
$\Lambda=$ 1.5 GeV, 
the central value in Ref.~\cite{Foroughi-Abari:2021zbm}.

We note that the PB process  
will also contribute to the dimuon final states, 
if one replaces $\chi$ with the muon in 
the Feynman diagram in Fig.~\ref{fig:pb-diagram}. 
We thus carry out an analysis in Appendix \ref{app sec: NA60} 
to compute the dimuon events arising from 
the PB process for the NA60 experiment, 
and find that they are consistent with the NA60 data \cite{NA60:2016nad}. 
We note that the collision conditions 
in the NA60 experiment are significantly different 
from the atmosphere, thus resulting in different predictions. 
Notably, both 
the high energy of the proton beam, 400 GeV, and 
the angular acceptance, $35<\theta<120$ mrad, of the NA60 experiment, 
play important roles in the proton bremsstrahlung process. 
As shown in Appendix \ref{app sec: NA60}, 
these two factors lead to a significantly suppressed 
splitting kernel when  
the off-shell form factor given in Eq.~\eqref{eq: off-shell FF} 
is taken into account.

\section{Earth attenuation}

To compute the MCP flux at underground 
neutrino detectors, one has to 
take into account the Earth attenuation 
effects. 
The energy loss of MCPs 
along a trajectory in Earth 
can be described by \cite{Gaisser:2016uoy, ArguellesDelgado:2021lek}
\be
- \frac{d E}{d X} =  
\varepsilon^2 (a + b E),
\label{eq: energy loss}
\ee
where $X$ is the slant depth traversed, 
and $a$ ($b$) is the parameter to describe energy loss due to 
ionization (radiation in scatterings 
with nuclei). 
We adopt the parameters for muons 
in the standard rock: $a = 0.233$ GeV/mwe
and $b = 4.64 \times 10^{-4}$ mwe$^{-1}$ 
with 1 mwe $= 100\ \rm{g/cm}^2$
\cite{Koehne:2013gpa}. 
The slant depth traversed by MCPs 
can be computed by 
\begin{equation*}
X = \rho \sqrt{ \left(R_e - d\right)^2 + R^2_e - 2 \left(R_e  - d\right) R_e \cos \left(\theta - \theta_{s}\right)},    
\end{equation*}
where $\rho = 2.6\ \rm{g/cm}^3$ is the density of
the standard rock, 
$d$ is the depth of the detector, 
$R_e$ is the radius of Earth, 
$\theta$ ($\theta_{s}$) 
is the zenith angle of the MCP viewed by 
the detector (the intersection point on the Earth's surface)
such that $\sin \theta_{s}/ (R_e -d)  = \sin \theta / R_e$ 
with $\theta_{s} < \pi/2$. 
The MCP flux with energy $E_\chi$  
at slant depth $X$ underground 
is related to the flux at the surface via \cite{Gaisser:2016uoy}
\be 
\frac{d^2 \Phi^D_\chi (X)}{d E_\chi d\Omega} = 
e^{\varepsilon^{2} b X} \frac{d^2 \Phi_\chi^s}
{d E^s_\chi d \Omega^s},
\ee
where 
$E_\chi^s =  (E_\chi + a/b)\ \exp\left(\varepsilon^2 b X\right) - a/b$.

\section{Signals at Super-K}

Large underground 
neutrino detectors are ideal places to detect 
light atmospheric MCPs \cite{Plestid:2020kdm, Kachelriess:2021man, ArguellesDelgado:2021lek}.
In our analysis we consider the Super-K experiment, a large water-Cherenkov detector
with a fiducial volume of 22.5 kton of water,
shielded by 1000 m rock  \cite{Super-Kamiokande:2002weg, Super-Kamiokande:2011lwo,
Super-Kamiokande:2021jaq}.
The dominant process to detect MCPs at Super-K 
is the elastic MCP-electron scattering  
\cite{ArguellesDelgado:2021lek} 
\begin{equation*}
\frac{d \sigma}{d E_{r}}=\varepsilon^{2} \alpha^{2} \pi 
\frac{ E_{r}+2 E_{\chi}^{2}/E_{r} - 
2 E_{\chi} - m_{e} - m_{\chi}^{2}/m_e } 
{E_{r} m_{e}\left(E_{\chi}^{2}-m_{\chi}^{2}\right)}, 
\label{eq: chi e scattering}
\end{equation*}
where 
$E_\chi$ is the energy of the incident MCP,
$E_r$ is the recoil energy of the scattered electron, 
and $m_e$ is the electron mass. 
The signal events can be obtained by
\be
S_i =
 2 \pi n_{e} {\cal E}  
\int d E_r f\left(E_{r}\right) 
\int d E_{\chi} 
\int d z
\frac{d^2 \Phi^D_\chi}{d E_{\chi} d z} 
\frac{d \sigma}{d E_{r}},
\ee
where $z=\cos\theta$, 
${\cal E}$ is the exposure, 
$2\pi$ is due to the integration over the spherical angle $\phi$, 
$n_{e}$ is the number of electrons 
per unit mass of water,  
and $f\left(E_{r}\right)$ is the detector efficiency. 

In our analysis we use the data with {a} total exposure of 359 kton-year, 
accumulated during the first four Super-K runs,  
with running time 1497 days, 
794 days, 562 days, and 2970 days,
respectively 
\cite{Super-Kamiokande:2011lwo,
Super-Kamiokande:2021jaq}.
We use the Super-K data with Cherenkov angles ranging from 38 to 50 degrees, 
which correspond to recoiled electrons. 
For the first three Super-K runs 
(total exposure {of} 176 kton-year), 
we use the efficiency curve given in figure 10 of
Ref.~\cite{Super-Kamiokande:2011lwo}. 
We bin the signal events using the same 
bins in the Super-K data \cite{Super-Kamiokande:2011lwo}: 
18 bins with a 4 MeV bin width for each 
in the recoil energy of 16-88 MeV.
For the Super-K-IV data (total exposure of 183 kton-year), 
we bin the signal events using the same 
bins in the Super-K data \cite{Super-Kamiokande:2021jaq}: 
31 bins with a 2 MeV bin width for each 
in the recoil energy of 16-78 MeV.

Because of the limited knowledge on the detailed background in the Super-K data, we use the background-agnostic method \cite{ArguellesDelgado:2021lek} 
to set the constraints.
In this method, for the bins where the number of data events is less than the expected MCP signal events, we use the Poisson distribution
to compute the likelihood for each data bin 
\begin{equation}
{\cal L}_i = \frac
{(S_i)^{D_i} \exp(-S_i)}{D_i !},
\label{eq: likelihood}
\end{equation}
where $D_i$ and $S_i$ 
are the data events
and the expected signal events 
in the $i$-th bin. Otherwise the likelihood
${\cal L}_i$ is taken to be 1. 
The total likelihood is  
${\cal L}= \Pi_i {\cal L}_i$, 
and the test {statistic} is  
\begin{equation}
    {\rm TS} = - 2 \log \left[ 
    \frac{{\cal L}(m_\chi, \varepsilon)}{{\cal L}(m_\chi,\varepsilon=0)} 
    \right].
\end{equation}
We use ${\rm TS} < 4.6$ to 
compute the $90\%$ confidence level limit. 
In the parameter space of interest, the majority of MCPs arrive at the Super-K detector from a zenith angle less than 90 degrees due to Earth attenuation effects. Consequently, the recoiled electrons in the signal process are likely to have a downward momentum. 
In contrast, the background events exhibit 
different angular distributions. 
In the low energy bins  
($16$ MeV $\lesssim E_r \lesssim 55$ MeV), 
the dominant backgrounds 
are due to electrons from low energy muon decays,  
and are expected to be isotropic \cite{Super-Kamiokande:2011lwo}. 
In the high energy bins 
($55$ MeV $\lesssim E_r \lesssim 88$ MeV),  
the dominant backgrounds are due to electron-neutrinos, 
and exhibit a higher probability of arriving at the detector 
from zenith angles exceeding 90 degrees 
compared to those at zenith angles less than 90 degrees 
\cite{Super-Kamiokande:2011lwo,
Super-Kamiokande:2021jaq}. 
We thus take advantage of this angular correlation between the incident MCP and the recoiled electron
by multiplying a factor of half in the likelihood function for the Super-K data events.

\section{Super-K constraints on atmospheric MCPs}

\begin{figure}[htbp]
\begin{centering}
\includegraphics[width= 0.5 \columnwidth]{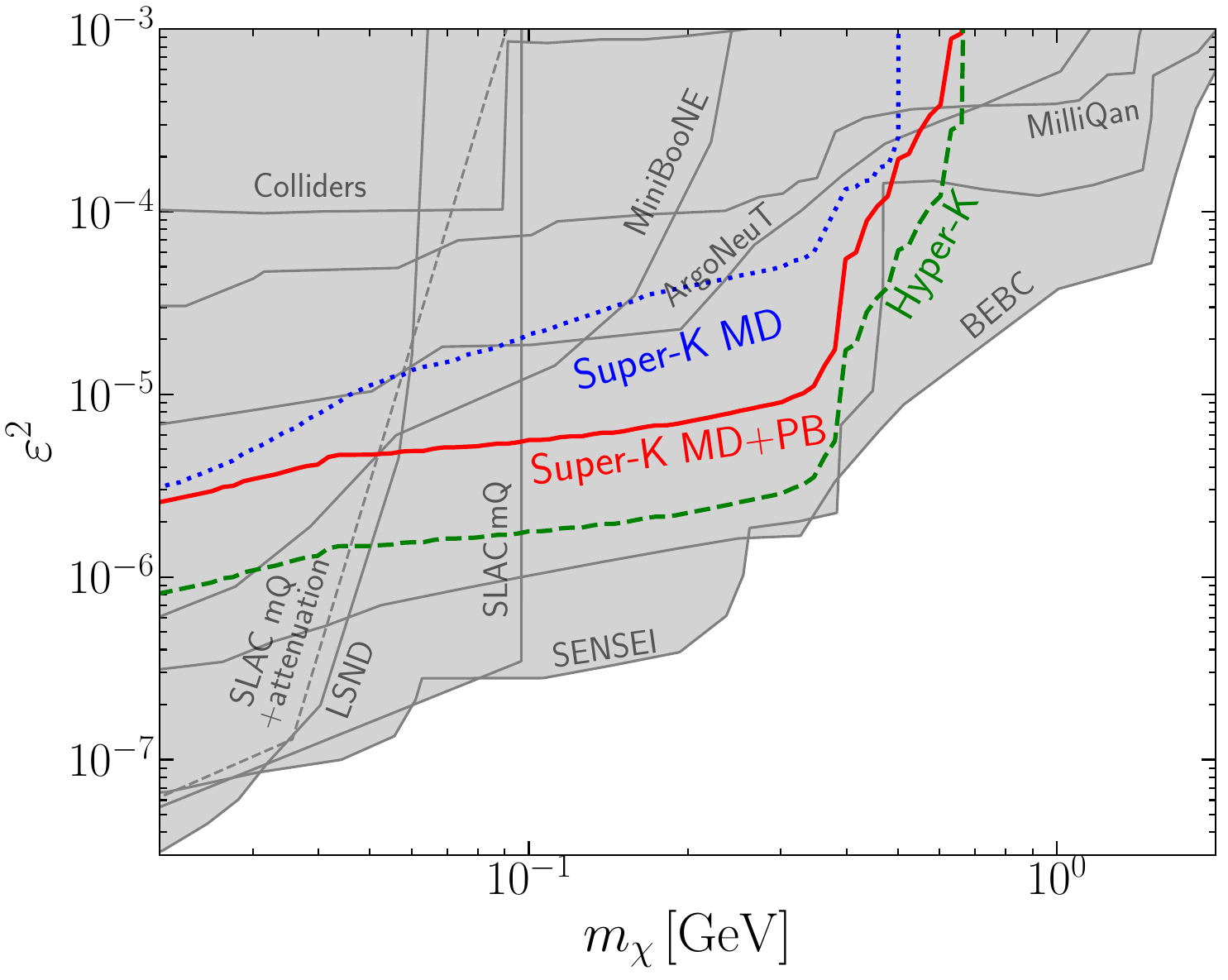}
\caption{Super-K constraints (I-IV data) on MCP 
at the $90\%$ CL 
from the MD process (blue dotted curve), 
and from both the PB and MD processes (red solid curve). 
{The} existing constraints are shown as 
shaded regions: 
ArgoNeuT \cite{ArgoNeuT:2019ckq},
MilliQan \cite{Ball:2020dnx},
Colliders \cite{Davidson:2000hf}, 
LSND \cite{Magill:2018tbb},
MiniBooNE \cite{Magill:2018tbb},
SENSEI \cite{SENSEI:2023gie},
BEBC \cite{Marocco:2020dqu},
SLAC mQ \cite{Prinz:1998ua},  
and the revised SLAC mQ 
(taking into account attenuation effects) 
\cite{Arefyeva:2022eba}. 
Hyper-K projections (green dashed curve) are computed 
assuming the same running time as the Super-K I-IV data.}
\label{fig:MCP_constrain_SuperK} 
\end{centering}
\end{figure}

Fig.~\ref{fig:MCP_constrain_SuperK} 
shows the new Super-K constraints on millicharged particles, 
by taking into account  
the proton bremsstrahlung process 
in the atmosphere. 
We compute the $90\%$ CL limits on {MCPs}  
from Super-K, both for the MD-only process
and for the combined contributions from the MD and PB processes.
Our MD-only limits agree with Refs.~\cite{Plestid:2020kdm,Kachelriess:2021man}, 
but not with Ref.~\cite{ArguellesDelgado:2021lek}; 
see Appendix \ref{appendix:MD} for the comparisons. 
After taking into account the contributions from 
the PB process, 
the Super-K limit on $\varepsilon^2$ 
is improved 
by $\sim$ 4 times
for an MCP mass larger than 0.1 GeV.
This improvement in the Super-K limit is largely because the number of MCP signals 
in the Super-K from the PB process 
is about $1 \sim 10$ times
larger than 
that from the MD process. 
Thus, the PB process 
is the dominant MCP production process 
for the sub-GeV MCP.

The Super-K limit on $\varepsilon^2$ computed in this work 
is shown to be better than the previously best experimental 
limits from the ArgoNeuT \cite{ArgoNeuT:2019ckq} 
by more than 4 times, 
in the mass range of (0.1-0.7) GeV. 
Recently, Ref.~\cite{Marocco:2020dqu} has obtained new limits on MCPs from the past experiment BEBC.
but which may have potential systematic uncertainties \cite{ArguellesDelgado:2021lek}.
After the first arXiv version of our analysis, 
SENSEI has released its new constraints on MCPs 
\cite{SENSEI:2023gie}. 
We further compute the sensitivities on 
atmospheric MCPs from Hyper-K, the successor to Super-K. 
In our analysis, we consider the fiducial 
volume of Hyper-K to be ten times of
Super-K, and we assume the same running time 
as the first four Super-K runs.

Recently, Ref.~\cite{Arefyeva:2022eba} reanalyzed the SLAC mQ limit by
taking into account the attenuation effects.
This opens up some previously
excluded parameter space for $m_\chi < 0.1$ GeV \cite{Arefyeva:2022eba}, 
as also shown in Fig.~\ref{fig:MCP_constrain_SuperK}. 
This also raises questions about the sensitivity of other
accelerator experiments in effectively detecting MCPs 
that are produced by a distant point source and need to penetrate rock or other materials to reach the detector. 
We note that angular deflections resulting from attenuation effects
play a less significant role in the total flux of atmospheric MCPs. 
Consequently, the Super-K limits on atmospheric MCPs,  as analyzed in this study, 
are more robust compared to accelerator limits in this particular regard.

We note that 
the sudden change in the upper bound on $\varepsilon^2$
near $m_{\chi} = m_{\rho}/2 \sim 0.38$ GeV 
is primarily due to the fact that the 
production rate of light MCPs 
in the atmosphere is significantly enhanced by 
the electromagnetic time-like form factor $F_V (k)$ 
of the $\rho/\omega$ 
vector meson; for MCP with mass exceeding  
$m_{\rho}/2$, the invariant mass of the 
off-shell photon, $\sqrt{k^2}$, is always larger than 
the mass of the $\rho/\omega$ meson 
so that the resonant production via the time-like 
form factor starts to diminish. 
We also note that perhaps there is a double counting 
between the enhancement via the $\rho/\omega$ form 
factor and the production of MCPs via 
$\rho/\omega \to \gamma^{*} \to  \chi \bar \chi$
in the MD process \cite{Chu:2020ysb}. 
However, since the PB process is much larger than the MD process 
for light MCPs in the MeV-GeV mass range, 
one {should} just consider the PB process and neglect 
the {vector} MD process to avoid the possible double-counting. 

We further note that
the study on the proton bremsstrahlung process can be extended to a variety of new physics particle
searches in atmospheric collisions and in low energy proton accelerators. 
For example, the proton bremsstrahlung process 
may also lead to improved limits for 
MCP searches in experiments with a proton beam, 
where only the meson decay process was analyzed previously.

\section{Conclusions}

In this work we study the PB process for the 
atmospheric MCP flux, which has not been considered 
in previous analyses, where only the MD process 
are included. 
To take into account the low-energy cosmic protons, 
we develop a new method to compute the PB process, 
instead of using the FWW approximation, 
where protons and the emitted photon are assumed to 
be relativistic. 
We find that the PB process is the dominant production channel 
for sub-GeV MCPs, 
which can give rise to a much larger MCP flux 
than the previously studied MD channel. 
The experimental limits from Super-K are 
improved by a factor of around 4 for $\varepsilon^2$, 
when the PB process is taken into account. 
This surpasses the previously best experimental 
limits from the ArgoNeuT \cite{ArgoNeuT:2019ckq} 
by half order of magnitude, 
in the mass range of (0.1-0.7) GeV. 
We note that the analysis presented here for the 
PB process can be extended 
to atmospheric productions of 
other light new particles in the hidden sector, 
and to MCP searches in terrestrial experiments 
with a low energy proton beam.

\begin{acknowledgments}

We thank 
Jinhan Liang, 
Wenxi Lu, 
Ligang Xia, 
Lan Yang, and 
Zicheng Ye 
for discussions.
The work is supported in part by the 
National Natural Science Foundation of China under Grant Nos.\ 12275128 
and 12147103.

\end{acknowledgments}

\appendix

\section{MCP signals from meson decays}
\label{appendix:MD}

In this section we provide a brief description of  
our calculation 
of MCPs produced in the MD process, 
and further compare it to those 
in Refs.~\cite{Plestid:2020kdm, Kachelriess:2021man, ArguellesDelgado:2021lek}. 
We find that our Super-K limits  
(when only the MD process is considered) 
is in ballpark agreement with Refs.~\cite{Plestid:2020kdm, Kachelriess:2021man}, 
but is about one order of magnitude weaker 
on $\varepsilon^2$ than Ref.~\cite{ArguellesDelgado:2021lek}.
{We summarize the main differences between 
our calculation and 
Ref.~\cite{ArguellesDelgado:2021lek} 
at the end of this section.}

The calculation for the MD process is the same as 
the PB process, except the energy spectrum of 
atmospheric MCPs 
given in Eq.~\eqref{eq: ground chi flux}.
For the MD process, one has
\be
 \frac{\mathrm{d}{N}^{\rm{MD}}_{ \chi}}{\mathrm{d} E^s_{\chi}} = 2
\sum_m
\int_1^{\infty} \mathrm{d} \gamma_m
\frac{d {N}_m (E_p)}{d \gamma_m }
F_m (E^s_{\chi}, \gamma_m),  
\label{eq: MD MCP spectra}
\ee
where  
$m$ denotes the parent meson in the decay process,
$\gamma_m = E_m/m_m$ is the meson boost factor,
$d {N}_m (E_p) / d \gamma_m $
is the spectra of the averaged multiplicity of mesons,
the factor 2 comes from 
the fact that
two MCPs are produced in each meson decay,
and 
$F_m (E^s_{\chi}, \gamma_m) $ is the MCP energy spectra 
in the lab frame, 
which is obtained by boosting the spectra in the rest 
frame of meson $m$ to the lab frame.
We use the EPOS model \cite{Pierog:2013ria}
in the CRMC package \cite{Ulrich:crmc} 
to compute $d {N}_m (E_p) / d \gamma_m $. 

\begin{figure*}[thbp]
\begin{centering} 
\includegraphics[width=0.4 \textwidth]{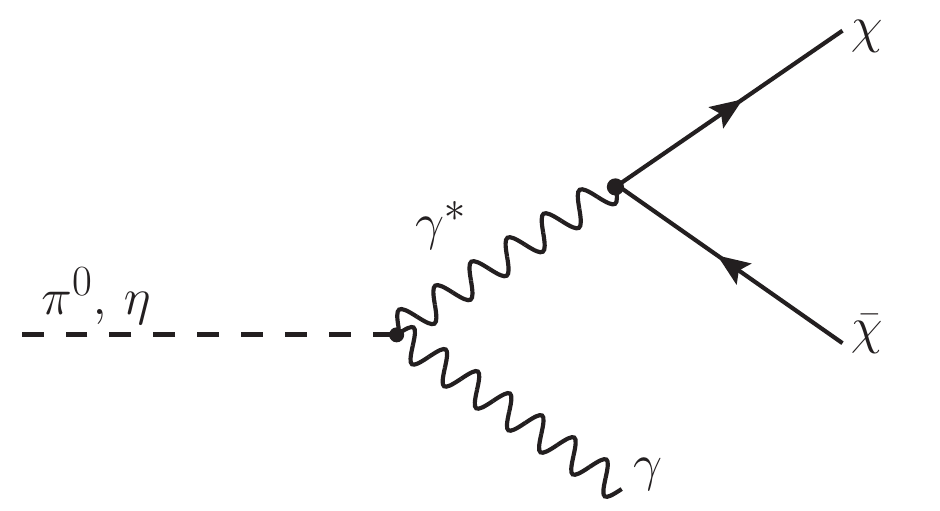}
\includegraphics[width=0.4 \textwidth]{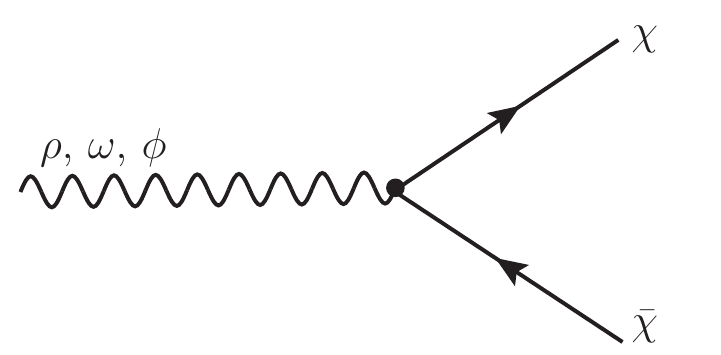}
\caption{Feynman diagrams for MCPs 
productions from meson decays 
in the atmosphere: pseudo-scalar meson decays (left), and
vector meson decays (right).} 
\label{fig:diagram:MD}
\end{centering}
\end{figure*}

There are two types of meson decays in the atmosphere 
in which MCPs can be produced:  
pseudo-scalar meson decays, 
and vector meson decays, 
as shown in Fig.~(\ref{fig:diagram:MD}).  
In our analysis, we consider both 
pseudo-scalar mesons ($\pi^0$ and $\eta$)
and vector mesons ($\rho$, $\omega$ and $\phi$). 
We assume that all the mesons decay promptly, 
since they have very short decay length
compared to the distance they travel to 
reach the Earth surface. 
For example, $\pi^0$ has the longest lifetime
$\tau_{\pi^0} = 8.52 \times 10^{-17}$ s
among the mesons we consider; 
the decay length of a GeV 
$\pi^0$ is only $0.2$ $\mu$m.

\begin{figure}[htbp]
\begin{centering}
\includegraphics[width= 0.5 \columnwidth]{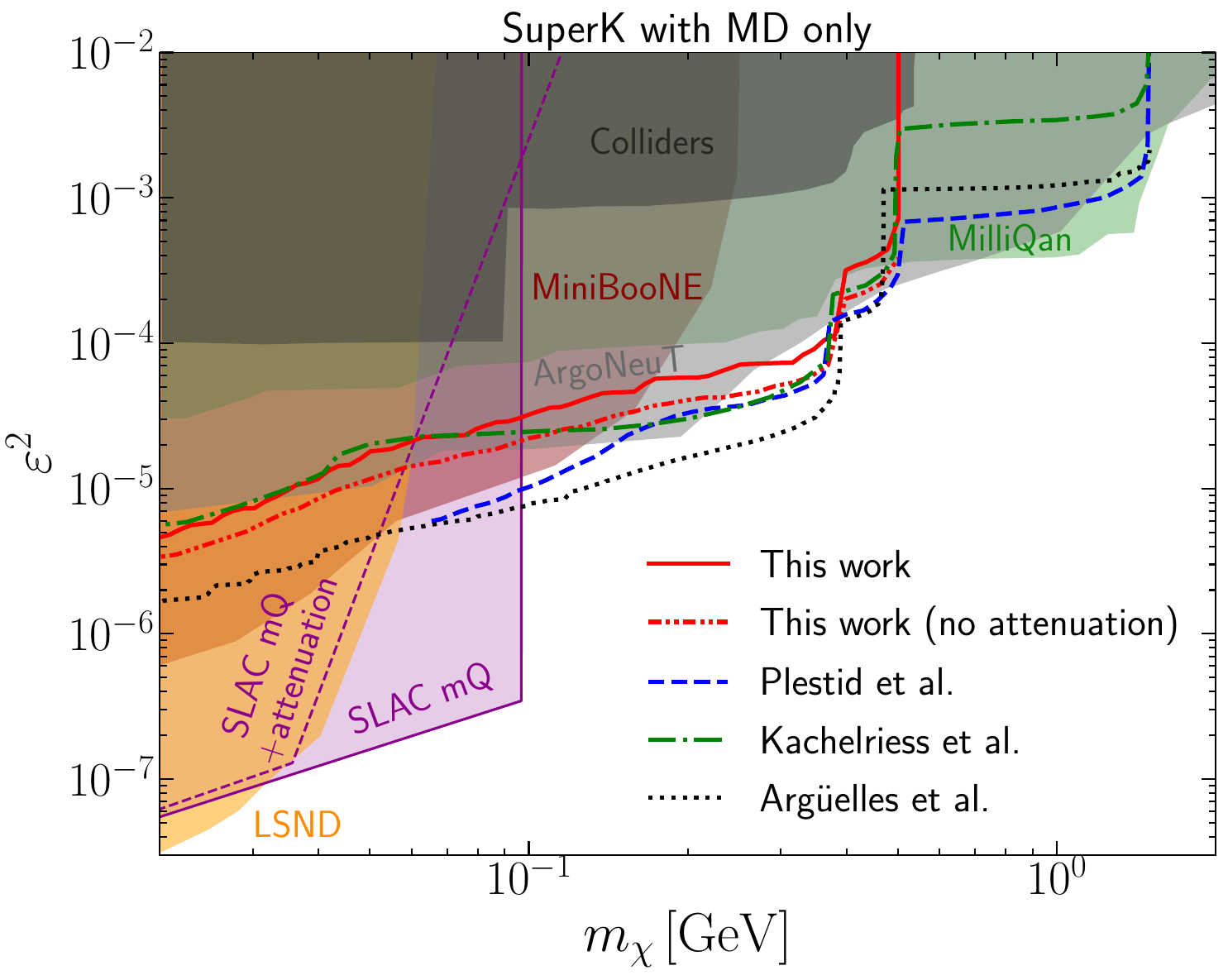}
\caption{Super-K $90\%$ CL constraints (data in the first three runs)
on MCPs with the DM channel only: 
our calculation (red solid curve), 
{our calculation without the Earth attenuation (red dash-dotted curve)}, 
Plestid et al.\ \cite{Plestid:2020kdm} (blue dashed curve),
Kachelriess et al.\ \cite{Kachelriess:2021man} (green dash-dotted curve), 
and
Arg\"uelles et al.\ \cite{ArguellesDelgado:2021lek} (black dotted curve). 
Existing constraints are also shown here as 
shaded regions: 
ArgoNeuT \cite{ArgoNeuT:2019ckq} (gray),
MilliQan \cite{Ball:2020dnx} (green),
Colliders \cite{Davidson:2000hf} (black), 
LSND \cite{Magill:2018tbb} (orange),
MiniBooNE \cite{Magill:2018tbb} (dark red),
SLAC {mQ} \cite{Prinz:1998ua} (purple-solid),  
and {the revised} SLAC mQ (with attenuation effects) \cite{Arefyeva:2022eba} (purple-dashed).
}
\label{fig:MCP_constrain_superk_md_compare} 
\end{centering}
\end{figure}

For the pseudo-scalar meson decay
$m \to \chi \bar\chi \gamma$,
one has 
\be
F_m (E_{\chi}, \gamma_m) 
= 
\int d x d y 
\frac{d^{2} {\rm Br} \left(m \to \chi \bar\chi \gamma \right)}{d x d y} 
\frac{\Theta\left(E_\chi- E_{-}\right) \Theta\left(E_{+} - E_\chi \right)}{E_{+} - E_{-}} ,
\label{eq: MCP spectra scalar}
\ee 
where 
$ x = m_{\chi \bar\chi}^2/m_{m}^2 $ 
with $m_{\chi \bar\chi}$ being the 
invariant mass of the final state 
$\chi \bar \chi$ pair, 
$ y = (E_\chi^r - E_{\bar\chi}^r) /E_\gamma^r $
with $E_f^r$ the energy 
of the final state particle $f$ in the 
rest frame of the meson, 
$E_{\pm} = \gamma_m (E_{\chi}^r 
\pm \beta_m p_{\chi}^r)$ 
with 
$p_\chi^r$ being
the MCP momentum  
in the rest frame of the meson, 
and 
${\rm Br} \left(m \to \chi \bar\chi \gamma \right)$ is the branching ratio. 
We use the Dalitz decay approach to compute 
the differential branching ratio \cite{Dalitz:1951aj}
\be
\frac{d^{2} {\rm Br} \left(m \to \chi \bar\chi \gamma \right)}{d x d y} 
=\frac{\varepsilon^2 \alpha}{\pi} 
{\rm Br}\left(m \rightarrow \gamma \gamma\right)
\mathcal{F}^2\left(m_{\chi \bar\chi}^{2}  \right) 
\frac{(1-x)^{3}}{4 x}\left(1+y^{2}+\frac{4 m_\chi^{2}}{m_{m}^{2} x}\right),
\label{eq: Dalitz}
\ee
where 
$\alpha$ is the fine structure constant,
${\rm Br}\left(\pi^0 \rightarrow \gamma \gamma\right) \simeq 0.99$, 
${\rm Br}\left(\eta \rightarrow \gamma \gamma\right) \simeq 0.39$ \cite{ParticleDataGroup:2020ssz},
and $\mathcal{F}\left(m_{\chi \bar\chi}^{2}  \right) $ 
is the meson form factor, which 
is given by
\cite{Berman:1960zz, Landsberg:1985gaz}
\bea
\mathcal{F} \left(m_{\chi \bar\chi}^{2}  \right) 
= 
\begin{cases}
1 + a_\pi \frac{m_{\chi \bar\chi}^{2}}{m_\pi^2} & \quad
\text{for $\pi^0$} \\
\frac{1}{1 -  m_{\chi \bar\chi}^{2}/ \Lambda_\eta^2} & \quad
\text{for $\eta$}
\end{cases},
\eea
where $a_\pi = 0.0335$ \cite{ParticleDataGroup:2018ovx}
and $\Lambda_\eta = 0.7191$ GeV \cite{NA60:2016nad}.

For the vector meson decay 
$m \to \chi \bar\chi$, one has 
\be
F_m (E_{\chi}, \gamma_m) 
= 
\frac{{\rm Br} \left(m \to \chi \bar\chi \right) }{E_{+} - E_{-}} 
\Theta\left(E_\chi- E_{-}\right) \Theta\left(E_{+} - E_\chi \right),
\label{eq: MCP spectra vector}
\ee 
where $E_{\pm} = \gamma_m \left(E_m /2 \pm \beta_m \sqrt{E_m^2/4 - m_\chi^2}\right)$.
The di-MCP branching ratio is computed
by rescaling the dilepton branching ratio
\cite{Landsberg:1985gaz}
\be
{\rm Br} \left( m \to \chi \bar\chi \right)
= \varepsilon^2 {\rm Br} \left( m \to \ell \bar\ell \right) 
\sqrt{\frac{1 - 4 m_\chi^2/m_m^2}{1 - 4 m_\ell^2/m_m^2}} 
\frac{1 + 2 m_\chi^2/m_m^2}{1 + 2 m_\ell^2/m_m^2},
\label{eq:meson:2body}
\ee
where %
${\rm Br} \left( m \to e^- e^+ \right)= 4.72 \times 10^{-5}$, 
$7.36 \times 10^{-5}$, and $2.973\times 10^{-4}$
for $\rho$, $\omega$, and $\phi$ respectively 
\cite{ParticleDataGroup:2020ssz}.

In Fig.~(\ref{fig:MCP_constrain_superk_md_compare}), we compare
our calculations in the MD channel 
with the results given in 
Refs.~\cite{Plestid:2020kdm, Kachelriess:2021man, ArguellesDelgado:2021lek}. 
We only consider the first three Super-K runs in Fig.~(\ref{fig:MCP_constrain_superk_md_compare}), 
because the Super-K-IV data were released after 
Refs.~\cite{Plestid:2020kdm, Kachelriess:2021man, ArguellesDelgado:2021lek}.
Note that the Earth attenuation effects are only considered in Ref.~\cite{ArguellesDelgado:2021lek}, 
but not in Refs.~\cite{Plestid:2020kdm, Kachelriess:2021man}.
When the Earth attenuation effects are neglected, 
our results are consistent with 
Refs.~\cite{Plestid:2020kdm, Kachelriess:2021man}. 
However, after taking into account the Earth attenuation effects, 
our results do not agree with Ref.~\cite{ArguellesDelgado:2021lek}. 
In fact, our limits are much weaker than
Ref.~\cite{ArguellesDelgado:2021lek}: 
For MCP mass below 0.4 GeV, 
the Super-K limit on $\varepsilon^2$ 
obtained in our analysis 
(with only the MD channel included) 
is higher than 
Ref.~\cite{ArguellesDelgado:2021lek} 
by a factor of $\sim$3. 
After a detailed comparison between our analysis in the MD channel  
and the analysis in Ref.~\cite{ArguellesDelgado:2021lek}, 
as well as the publicly 
available codes \cite{Arguelluesdata} provided by authors of 
Ref.~\cite{ArguellesDelgado:2021lek}, we find a number of 
differences between the two analyses, which are summarized below.

\begin{enumerate}

\item

We noticed an error in the
branching ratios (BR) in Ref.~\cite{ArguellesDelgado:2021lek}: 
The MCP BR from  meson decays in Ref.~\cite{ArguellesDelgado:2021lek} are 
a factor of 2 larger than 
Eq.~\eqref{eq: Dalitz} and Eq.~\eqref{eq:meson:2body}.

\item 

We noticed an error in computing the number of electrons $n_e$ in Super-K in \cite{Arguelluesdata}.
The ratio between the number of electrons 
and the number of H$_2$O molecules in water was taken 
to be 18 in \cite{Arguelluesdata}. In our analysis, we use 10.  

\item

The method to compute the Earth attenuation 
in our analysis is different from 
Ref.~\cite{ArguellesDelgado:2021lek}. 
By using the same meson flux data and comparing the final results, 
we have inferred that the Earth attenuation effects 
considered in Ref.~\cite{ArguellesDelgado:2021lek} 
result in a signal that is twice the magnitude of our analysis.

\item

In our calculations, we use the 
one-dimensional approximation,  
which leads to an isotropic MCP flux at the 
surface of Earth for $\theta_s < \pi/2$.
Ref.~\cite{ArguellesDelgado:2021lek} 
used a different method in which 
the MCP flux is found to be not isotropic 
at the surface of Earth. However, 
no significant effects on the final results 
were observed due to the differences in angular distributions 
of the MCP flux.

\end{enumerate}

Taken together, the aforementioned four differences 
result in a factor of $\sim$8 discrepancy between 
Ref.~\cite{ArguellesDelgado:2021lek} and our analysis.
This is consistent with the limits on $\varepsilon^2$ given by Ref.~\cite{ArguellesDelgado:2021lek}, 
which are lower than ours by a factor of $\sim$3, 
since the MCP signal in Super-K is 
approximately proportional to $\varepsilon^4$.

\section{Our new PB method versus FWW}
\label{appendix: our FWW compare}

In this section, 
we compare the atmospheric MCP flux arising in the PB process 
that are obtained in the following two different methods:
(1) our new method (denoted as ``New PB''), 
and 
(2) the FWW approximation 
\cite{Fermi:1924, Williams:1934, Weizsacker:1934} (denoted as ``FWW''). 
We also compare the Super-K constraints 
obtained with two different cosmic ray spectra: 
(1) the power-law spectrum in Eq.~\eqref{eq: cosmic proton spectrum at top}
(denoted as ``PL''), 
and 
(2) the actual cosmic ray data as given in PDG \cite{ParticleDataGroup:2018ovx} 
(denoted as ``PDG'').

The splitting kernel in the FWW approximation
is given by \cite{Kim:1973he, Tsai:1973py, Blumlein:2013cua}
\be
\frac{d^2 \mathcal{P}^{\rm FWW}_{p \to \gamma^* p}}
{d E_{k} d \cos\theta_{k}} = 
\left|\mathbf{J}(z , p^2_T)\right|
\frac{d^2 \mathcal{P}^{\rm FWW}_{p \to \gamma^* p}}{d z d p_T^2} =
\left|\mathbf{J}(z , p^2_T)\right|
\omega (z, p_T^2) ,
\label{eq: PB FWW}
\ee
where {
$k^\mu = (E_k, \vec k)$ is the 4-momentum of the virtual photon,
$\theta_k$ is the angle between the virtual 
photon and the initial proton, 
$p_T = |\vec{k}|\sin\theta_k$ is the transverse 
momentum, 
$z =  \cos\theta_k |\vec{k}|/|\vec{p}_p|$ with 
$\vec{p}_p$ being the momentum of 
the initial proton,
$\left|\mathbf{J}(z , p^2_T)\right|$ is
the determinant of the Jacobian matrix
between $(z, p_T^2)$ and $(E_k, \cos\theta_k)$, }
and $\omega(z, p_T^2)$ is given by 
\cite{Kim:1973he, Tsai:1973py, Blumlein:2013cua} 
\be
\begin{aligned}
\omega\left(z, p_{T}^{2}\right)& \simeq & \frac{\alpha}{2 \pi H}\left\{\frac{1+(1-z)^{2}}{z}-2 z(1-z)\right.
 \left(\frac{2 m_{p}^{2}+k^{2}}{H}-z^{2} \frac{2 m_{p}^{4}}{H^{2}}\right) \\
&& + \,\; 2 z(1-z)\left(z+(1-z)^{2}\right) \frac{m_{p}^{2} k^{2}}{H^{2}} \left. + 2 z(1-z)^{2} \frac{k^{4}}{H^{2}}\right\}, 
\end{aligned}
\ee
where $H=p_{T}^{2}+(1-z) k^{2}+z^{2} m_{p}^{2}$. We note that the FWW approximation is valid
in the relativistic and collinear limit: 
$E_p, E_k, E_{p'} \gg m_p, \sqrt{k^2}, p_{T}$
\cite{
Kim:1973he, Tsai:1973py, Blumlein:2013cua,
deNiverville:2016rqh,
Feng:2017uoz,
Tsai:2019buq,
Foroughi-Abari:2021zbm,
Du:2021cmt}. 
Thus, in our analysis, we impose 
in the lab frame 
the following three conditions 
{to gurantee the validity of} 
the FWW approximation 
(denoted as the ``FWW cuts''): 
\begin{enumerate}

    \item $p_T < 0.1\ E_k$ \cite{Blumlein:2013cua, deNiverville:2016rqh},  
    
    \item $p_T < 1$ GeV 
\cite{Blumlein:2013cua, deNiverville:2016rqh}, 

    \item 
$\left|q_{\min }^{2}\right| < \Lambda^2_{\mathrm{QCD}}$
\cite{ Feng:2017uoz},
where
$
\left|q_{\min }^{2}\right| \approx \left[p_{T}^{2}+(1-z) k^{2}+z^{2} m_{p}^{2}\right]^{2}/
\left[{4 E_{p}^{2} z^{2}(1-z)^{2}}\right]
$ 
\cite{Kim:1973he, Tsai:1973py}
and $\Lambda_{\mathrm{QCD}} \simeq 0.25$ GeV is the QCD scale.
\end{enumerate}

\begin{figure}[htbp]
\begin{centering}
\includegraphics[width= 0.45 \columnwidth]{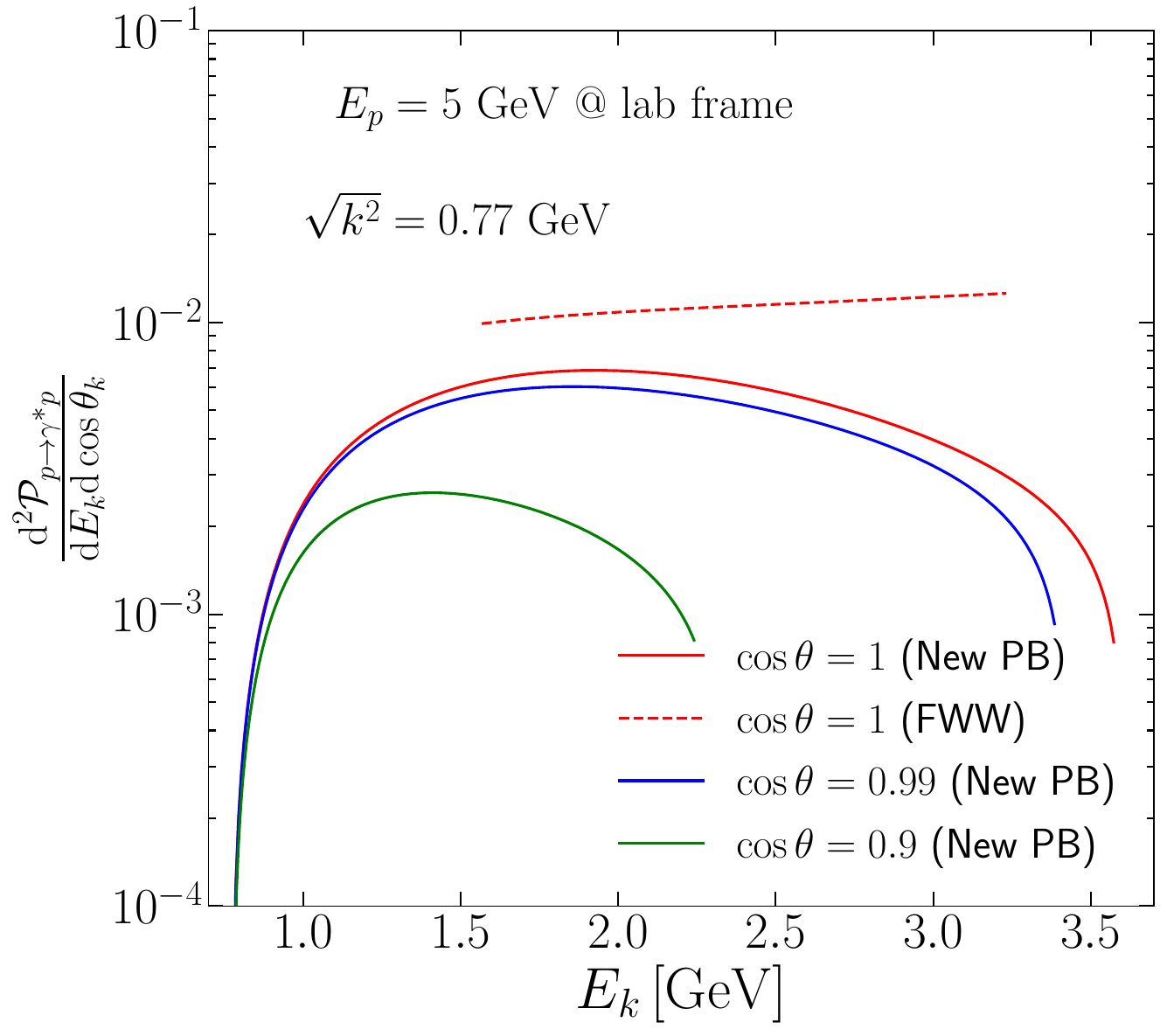}
\includegraphics[width= 0.45 \columnwidth]{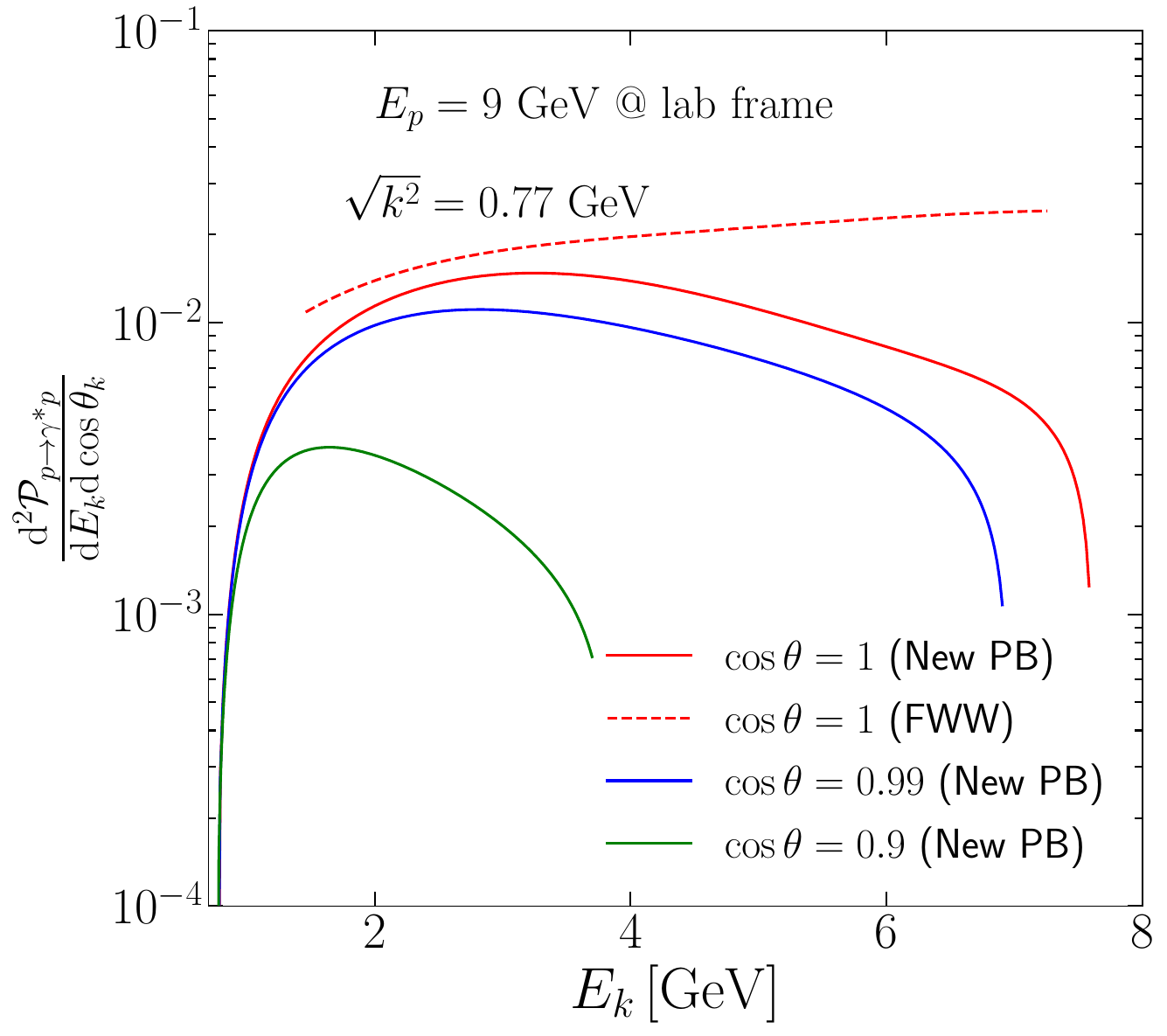}
\caption{The splitting kernel (without the proton
form factor) in the lab frame 
computed in our method 
and in the FWW approximation, 
for proton energy $E_p=5$ GeV (left), 
and for $E_p=9$ GeV (right). 
Three different angles in our method 
are plotted: 
$\cos\theta=1$ (red-solid), 
$\cos\theta=0.99$ (blue-solid),
and 
$\cos\theta=0.9$ (green-solid). 
In the FWW approximation,  
we impose the ``FWW cuts'',
under which the splitting kernel 
is zero for both 
$\cos\theta=0.99$
and $\cos\theta=0.9$, and thus only 
$\cos\theta=1$ (red-dashed) is shown.}
\label{fig:splitting:kernel:compare} 
\end{centering}
\end{figure}

In Fig.~(\ref{fig:splitting:kernel:compare}), 
we compare the splitting kernel computed in 
our new method with that in the FWW approximation. 
We consider two proton energies in Fig.~(\ref{fig:splitting:kernel:compare}): 
$E_p=5$ GeV and 9 GeV, as the low energy cosmic 
proton has a larger flux. 
The splitting kernels 
in the $\cos\theta_{k} = 1$ case (the forward direction) 
in the FWW approximation are larger than 
those in our new method; 
however, the soft and hard photons in the $\cos\theta_{k} = 1$ case 
are rejected by the ``FWW cuts''.
Furthermore, the splitting kernels 
in the FWW approximation
vanish for $\cos\theta_{k} = 0.99$ and $0.9$, 
after imposing the ``FWW cuts''. 
Thus, we conclude that the FWW approximation
is not suitable for the PB process 
in the atmosphere for the protons 
energy below $\sim 10$ GeV.

\begin{figure}[htbp]
\begin{centering}
\includegraphics[width= 0.5 \columnwidth]{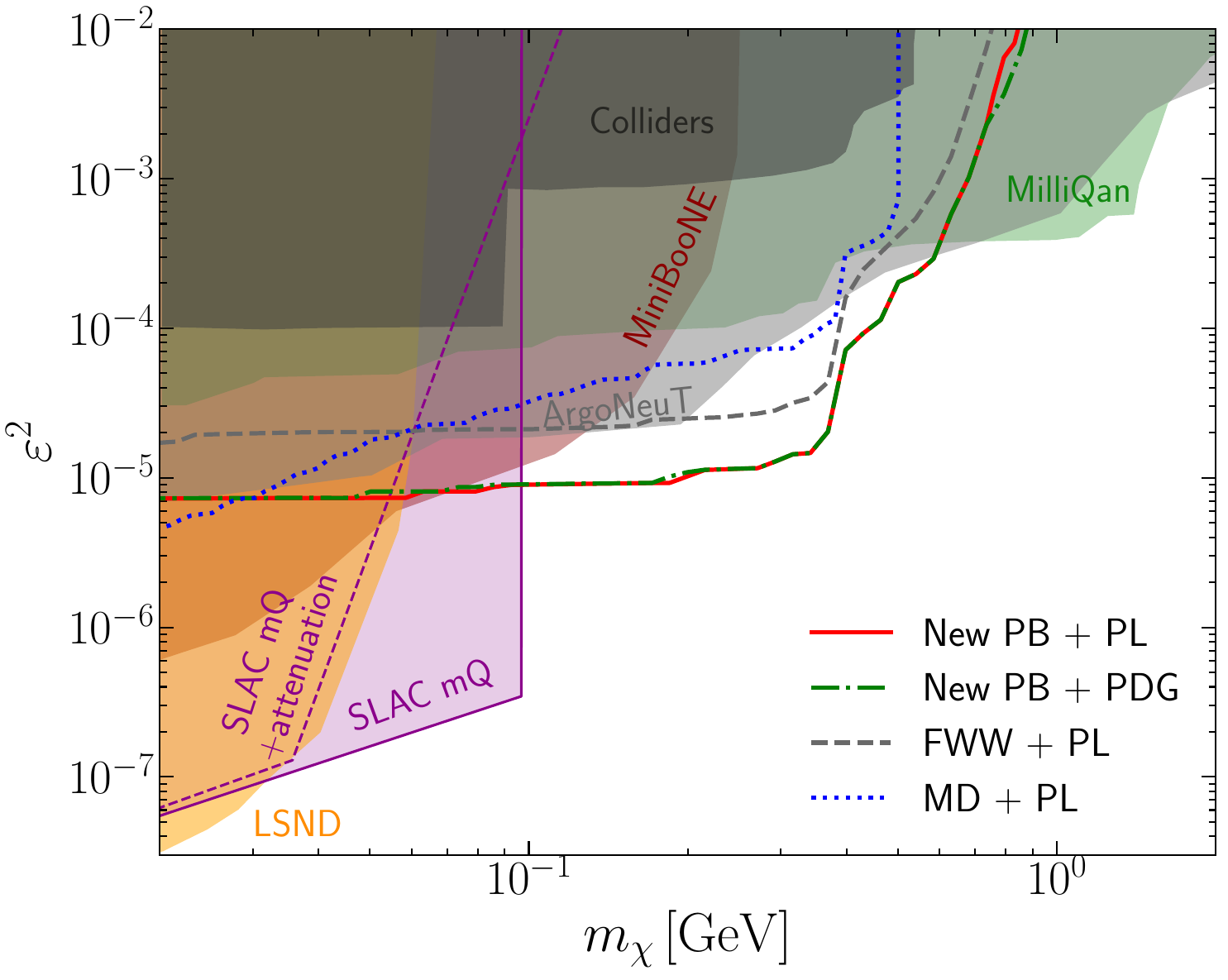}
\caption{Super-K $90\%$ CL limits 
(data in the first three runs)
on MCPs with the PB process:
{(1)} our new PB method  
with the power law cosmic ray model in 
Eq.~\eqref{eq: cosmic proton spectrum at top}
(red solid curve),
{(2)} the FWW approximation with 
the power law cosmic ray model (gray dashed curve), 
and 
{(3)} our new PB method  
with the cosmic ray data in Ref.~\cite{ParticleDataGroup:2018ovx}  (green dash-dotted curve).  
The MD process with the power law cosmic ray model 
(blue dotted curve) is also shown. 
Existing constraints are also shown here as 
shaded regions: 
ArgoNeuT \cite{ArgoNeuT:2019ckq} (gray),
MilliQan \cite{Ball:2020dnx} (green),
Colliders \cite{Davidson:2000hf} (black), 
LSND \cite{Magill:2018tbb} (orange),
MiniBooNE \cite{Magill:2018tbb} (dark red),
SLAC mQ \cite{Prinz:1998ua} (purple-solid), 
and {the revised} SLAC mQ (with attenuation effects) \cite{Arefyeva:2022eba} (purple-dashed).}
\label{fig:MCP_constrain_superk_check_1} 
\end{centering}
\end{figure}

In Fig.~(\ref{fig:MCP_constrain_superk_check_1}), 
we compare the Super-K constraints 
(data in the first three runs) 
obtained in our new method 
and in the FWW approximation. 
To speed up the calculation in the FWW approximation,
we take the average energy $E_\chi = E_k/2$ for MCPs.
The upper bound on $\varepsilon^2$ from Super-K 
computed in our new method is a factor of $\sim$2 better 
than that in the FWW method 
(with the ``FWW cuts'' taken into account), 
for MCPs in the mass range of $0.02$ GeV $\lesssim m_\chi \lesssim 0.35$ GeV, 
where the power law spectrum is used for the cosmic protons. 
We note that the weaker limits in the FWW method is due to its failure to account for a significant portion of low-energy protons, which constitute the dominant component of cosmic rays. 
As a result, the FWW approximation significantly underestimates the MCP flux.
We further note that the Super-K limits from the PB channel 
are significantly better than the MD channel, 
both in the FWW method and in our new method,
indicating that the PB channel 
is the dominant production channel of sub-GeV atmospheric MCPs.

The Super-K limits from the two different profiles of 
cosmic protons, ``PL'' and ``PDG'', are almost identical, 
where our new PB method is used, 
as shown in Fig.~(\ref{fig:MCP_constrain_superk_check_1}). 
Thus we mainly use the PL profile in our analysis. 
We note that 
although the amount of the cosmic protons
in the PL profile at very low energy 
can be significantly
larger than the actual cosmic ray data, 
these very low energy protons do not provide
the leading contributions to the PB process, 
as the limits from the two profiles are in 
excellent agreement with each other.
This is because protons with energy $\lesssim 2.6$ GeV 
in the lab frame 
are unable to radiate off-shell photons with $\sqrt{k^2}$ 
near the masses of the $\rho/\omega$ mesons. 
For that reason, 
the MCP production from these very low energy protons 
cannot occur via the resonances of the 
time-like form factors of the $\rho/\omega$ vector mesons.
We note that this is also in agreement with the sudden change in the limit 
near $m_{\chi} = m_{\rho}/2$.

\section{Comparison with NA60 data}
\label{app sec: NA60}

In this section we compute the mass spectrum of the di-muon from the PB process 
in the NA60 experiment \cite{NA60:2016nad}, 
which is a fixed target experiment with a 400 GeV proton beam
and angular acceptance ranging from
35 to 120 mrad.
We find that our result are consistent
with the mass spectrum measured by the NA60 apparatus~\cite{NA60:2016nad}.

The di-muon mass spectrum in NA60 for PB process can be computed as
\be
\begin{aligned}
\frac{d N}{d m_k} = &\sum_I  N_p  n_a \frac{\rho_I}{A_I} z_I \sigma_I \frac{e^2}{6 \pi^2}
\frac{1}{m_k} 
\sqrt{1 - \frac{4 m_\mu^2}{m_k^2}}
\left( 1 + \frac{2 m_\mu^2}{m_k^2} \right)\\
& \times \left| F_V(m_k)\right|^2
\int G\left( \tan\theta_k, m_k\right) d \tan\theta_k.
\end{aligned}
\ee
Here $m_k$ is the invariant mass of 
the muon pair.
$N_p$ is the total number of protons on targets, taken as $N_p =  7725 \times 2 \times 10^9$ for NA60~\cite{uras2007study, uras2011low}.
$n_a$ is the Avogadro's constant.
The subscript ``$I$" represents the material of the NA60 target, including Al, U, W, Cu, In, Be, and Pb.
$\rho_I$ is the mass density of the material $I$. 
$A_I$ is the atomic weight of nucleons in the target nucleus $I$,
$z_I$ is the thickness of the target material $I$~\cite{uras2007study, uras2011low}.
$\sigma_I$ is the cross section between the proton and nucleus $I$.
$m_\mu$ is the muon mass.
$F_V(m_k)$ is the VMD form factor in Eq.~\eqref{eq:PB_formfactor}.
The integration region is taken from tangent 35 to 120 mrads \cite{NA60:2016nad}.
The cross section $\sigma_I$ is given by~\cite{Gondolo:1995fq}:
\be
\sigma_I = A_I^{2/3} \sigma_{pN},
\ee 
where $\sigma_{pN} \approx 10$ mb. 
The $G$ function is defined as
\be
G(\tan\theta_k,m_k) \equiv
\int d E_{k}
\left|\frac{\vec{k}}{\vec{k}^0}\right|\left|
\frac{\tan\theta_k}{\left(1+\tan^2\theta_k\right)^{3/2}}\right|
\left|F_* \left(p_p-k\right)\right|^2
P_{\theta}(E_k)
\frac{d^2 \mathcal{P}_{p \to \gamma^* p}}{d E^0_{k} d \cos\theta_{k}^0},
\label{eq:spliiting_to_tan}
\ee
where $|{\vec{k}}/{\vec{k}^0}|\times
|{\tan\theta_k}/{\left(1+\tan^2\theta_k\right)^{3/2}}|$ 
is the Jacobian between $\left(E^0_{k}, \cos\theta_{k}^0\right)$ in the CM frame 
and $(E_k, \tan\theta_k)$ in the lab frame, 
$P_{\theta}(E_k)$ is the probability that 
both $\mu^+$ and $\mu^-$ 
from the virtual photon are 
in the angular acceptance of NA60, 
which is 35-120 mrads. 
The integral over $E_{k}$ in 
Eq.~\eqref{eq:spliiting_to_tan} 
is non-zero when the following condition is satisfied 
\begin{equation}
m_k^2 (1+\gamma^2_p\tan^2\theta_k) < 
E_{\rm th}^2 (1+\tan^2\theta_k), 
\end{equation}
where $\gamma_p = (1-\beta_p^2)^{-1/2}= E^0_p/m_p$ with 
$E^0_p$ being the energy of the initial proton in the CM frame, and 
$E_{\rm th}=\left(4 (E^0_p)^2 - 4 m^2_p + m^2_k\right)/4 E^0_p$ 
is the maximal energy of the virtual photon in the CM frame. 
In this case, 
the lower limit of $E_{k}$ is $E_{k}^{-}$ or $m_k$, 
whichever is larger, 
and the upper limit of $E_{k}$ is $E_{k}^{+}$, 
where 
\begin{equation}
E^{\pm}_k =
\frac{\gamma_p E_{\rm th} (1+\tan^2\theta_k) \pm \gamma_p \beta_p
\sqrt{E^2_{\rm th}(1+\tan^2\theta_k) 
-m^2_k(1 + \gamma_p^2 \tan^2\theta_k)}}{1 + \gamma_p^2 \tan^2\theta_k}.    
\end{equation}
To facilitate the calculation, for the $P_{\theta}(E_k)$ function, 
we have used the probability for the virtual photon with $\theta_k=77.5$ mrad 
for all virtual photons;
we thus require the opening angle of the di-muon 
pair in the lab frame to be smaller than 85 mrads. 
We note that $P_{\theta}(E_k)$ is smaller 
for virtual photons with other angles, 
making our analysis an optimistic estimation.

\begin{figure}[htbp]
\begin{centering}
\includegraphics[width= 0.5 \columnwidth]{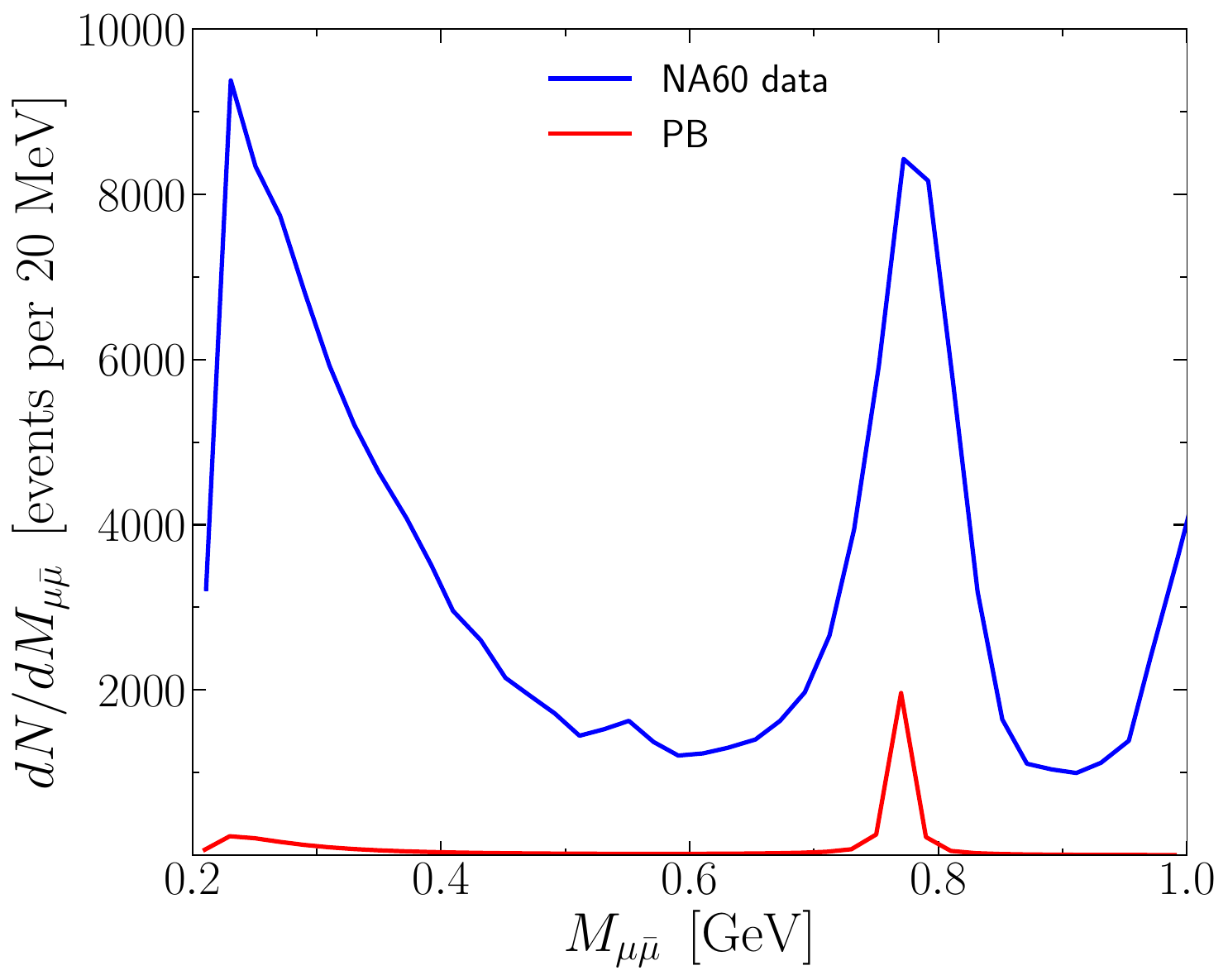}
\caption{Di-muon events as a function of the invariant mass 
in the NA60 experiment: 
(1) NA60 data (blue), 
(2) predictions from the PB process (red).}
\label{fig:compare_NA60_signal} 
\end{centering}
\end{figure}

Fig.~\ref{fig:compare_NA60_signal}
compares the di-muon events measured by
NA60 with those from the PB process.
We find that the di-muon events from the PB process are   
at least one order of magnitude
smaller than the NA60 data, 
except for the bin near the resonance at 0.77 GeV, 
where the PB process 
is $\sim$4 times smaller than the NA60 data. 
In our analysis, we have assumed a $100 \%$ efficiency. 
Thus, the di-muon events from the PB process 
should be even smaller than those in Fig.~\ref{fig:compare_NA60_signal}, 
when the experimental efficiency is taken into account.  
In conclusion, we find that the di-muon events 
from the PB process is 
consistent with the NA60 data \cite{NA60:2016nad}.

\begin{figure}[htbp]
\begin{centering}
\includegraphics[width= 0.45 \columnwidth]{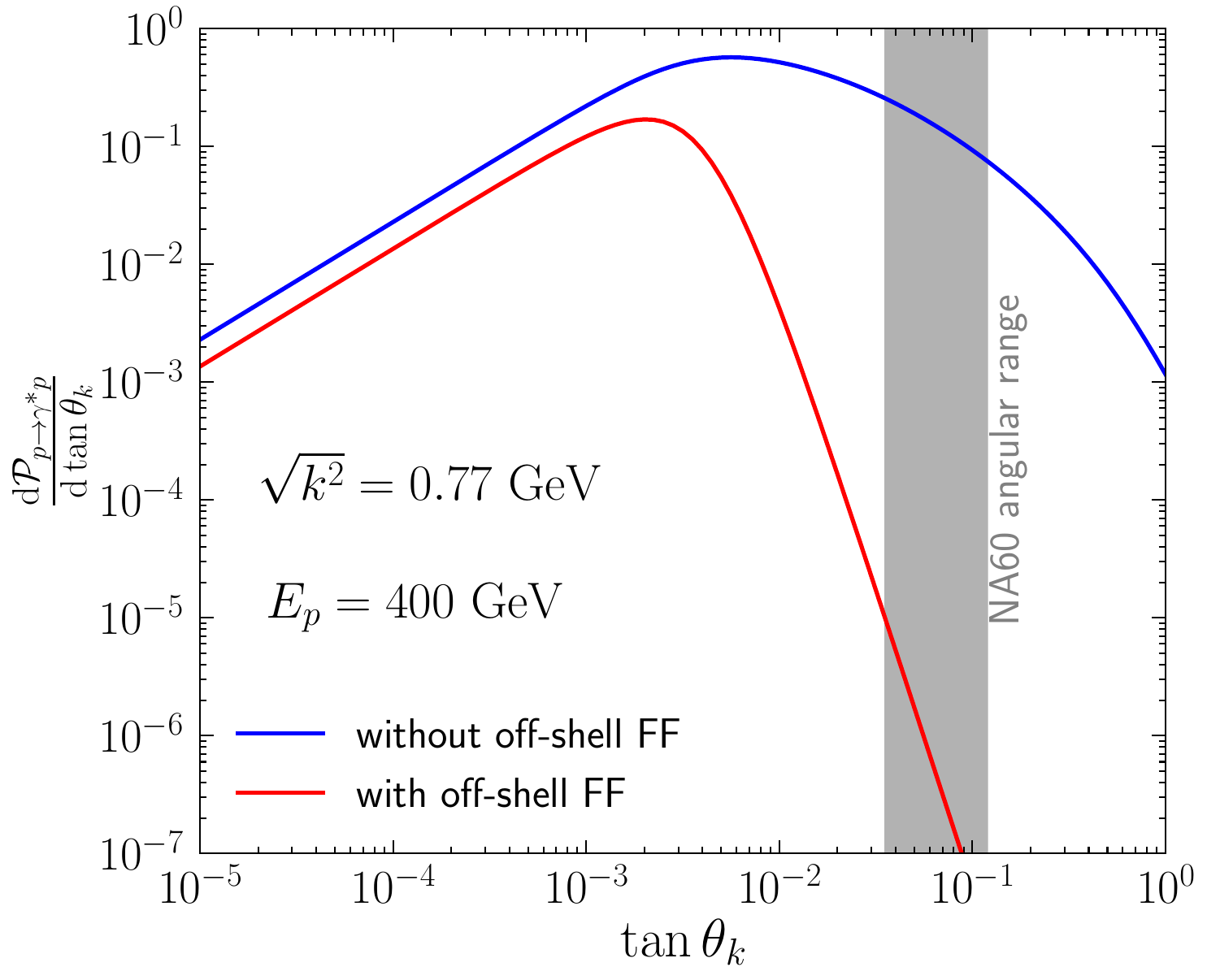}
\includegraphics[width= 0.45 \columnwidth]{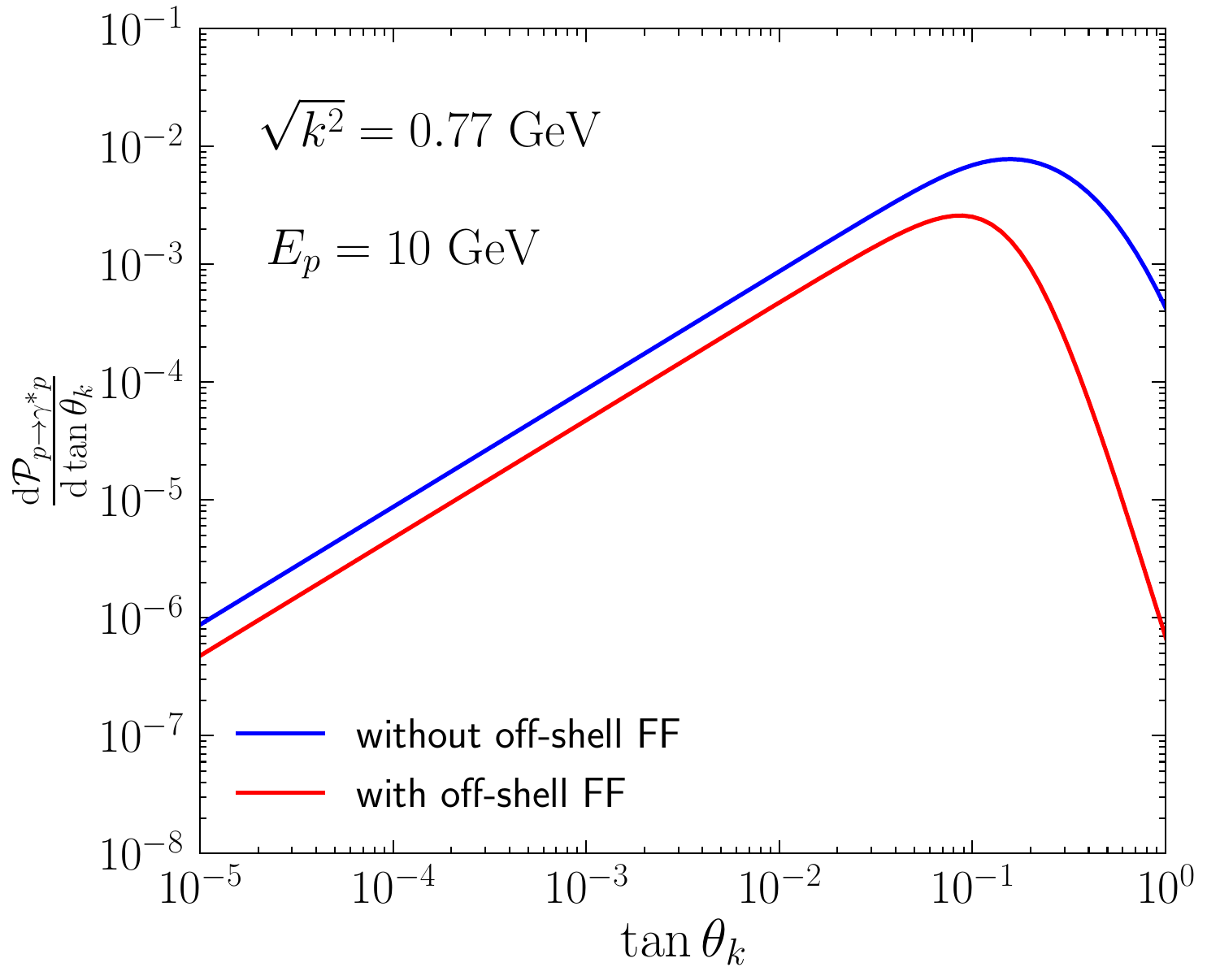}
\caption{The splitting kernel in the lab frame 
for the proton energy at 400 GeV (left panel), 
which is the proton beam energy at NA60, 
and 10 GeV (right panel), 
which is the characteristic energy scale of cosmic ray protons in the atmosphere. 
The splitting kernels computed with (without) the off-shell form factor 
are labeled by red (blue) curves. 
The gray shaded band in the left panel 
indicates the angular acceptance range of NA60, 
which is 35-120 mrad.}
\label{fig:compare_spliiting_off_FF} 
\end{centering}
\end{figure}

To provide a more comprehensive understanding of the results, 
we compute the splitting kernel in the lab frame:
\be
\frac{d P_{p\to p\gamma^*}}{ d\tan \theta_k} \equiv
\int d E_{k}
\left|\frac{\vec{k}}{\vec{k}^0}\right|\left|
\frac{\tan\theta_k}{\left(1+\tan^2\theta_k\right)^{3/2}}\right|
\left|F_* \left(p_p-k\right)\right|^2
\frac{d^2 \mathcal{P}_{p \to \gamma^* p}}{d E^0_{k} d \cos\theta_{k}^0}. 
\label{eq:splitting:kernel:lab}
\ee
Note that 
Eq.~\eqref{eq:splitting:kernel:lab} 
is the same as 
Eq.~\eqref{eq:spliiting_to_tan} 
except for the $P_\theta$ function.  
In 
Fig.~\ref{fig:compare_spliiting_off_FF} 
we compare the splitting kernel computed via Eq.~\eqref{eq:splitting:kernel:lab} 
against that computed without the off-shell form factor  
(i.e., setting $F_* (p'^2)=1$ in Eq.~\eqref{eq:splitting:kernel:lab}). 
Because the PB process has a significant contribution to 
the di-muon events close to the $\rho/\omega$ resonance, 
as shown in 
Fig.~\ref{fig:compare_NA60_signal}, 
we compute the splitting kernels with 
$\sqrt{k^2}=0.77$ GeV in 
Fig.~\ref{fig:compare_spliiting_off_FF}. 
The left-panel figure of 
Fig.~\ref{fig:compare_spliiting_off_FF} 
shows the splitting kernels 
at the proton beam energy at NA60, 
$E_p = 400$ GeV. 
We find that the inclusion of the off-shell form factor 
significantly reduces the splitting kernel 
at the NA60 beam energy  
for $\tan\theta_k \gtrsim 3 \times 10^{-2}$. 
In particular, 
in the angular acceptance of the NA60 experiment, 
$35 < \theta < 120$ mrad, 
the inclusion of the off-shell form factor 
reduces the splitting kernel by more than 
four orders of magnitude. 
In the right-panel figure of 
Fig.~\ref{fig:compare_spliiting_off_FF}, 
we also compute the splitting kernels 
at 10 GeV, 
the characteristic energy scale of cosmic ray protons in the atmosphere. 
We find that for 10 GeV protons the inclusion of the off-shell form factor 
also reduces the splitting kernel for large 
polar angle, $\tan\theta_k\gtrsim 10^{-1}$. 
Comparing the two panel figures in 
Fig.~\ref{fig:compare_spliiting_off_FF}, 
we find that the off-shell form factor leads 
to a much more significant suppression for 
substantial polar angles 
at the NA60 energy of 400 GeV than 10 GeV. 
We also note that in the very forward direction  
($\theta_k \lesssim 10^{-3}$), 
the inclusion of  the off-shell form factor plays 
a negligible role for both $E_p=400$ GeV and $E_p=10$ GeV. 
This explains why the PB process is crucial for the 
MCP production in atmospheric collisions 
but remains insignificant in the NA60 experiment.

\normalem
\bibliography{ref.bib}{}
\bibliographystyle{utphys28mod}

\end{document}